\documentclass[aip,pop,sd,reprint,graphicx]{revtex4-1}
\usepackage{amssymb, amsmath,framed,amsthm}

\makeatletter
\@ifpackageloaded{stix} % if 'stix' packages is loaded
{
	\newcommand{\bm}{\boldsymbol}
}{                      % if not loaded
  \renewcommand{\bm}[1]{{\boldsymbol {\mathrm #1}}} % non-italic bold
	\usepackage{mathrsfs}
	\usepackage{txfonts}              % Times-based fonts for text and math (equations become tight)
}
\makeatother

\usepackage[unicode=true,pdfusetitle,bookmarks=true,bookmarksnumbered=false,bookmarksopen=false,breaklinks=false,pdfborder={0 0 0},backref=false,colorlinks=true]{hyperref}
\hypersetup{citecolor=blue,filecolor=blue,linkcolor=blue,urlcolor=blue}
\usepackage{color,soul}
\usepackage{siunitx} % Comprehensive (SI) units
\usepackage{graphicx}% Include figure files
\usepackage{dcolumn}% Align table columns on decimal point
\usepackage{cancel}
\usepackage{diagbox} 
\usepackage{multirow}

\newcommand{\p}{\partial}
\newcommand{\f}[2]{\frac{#1}{#2}}
\newcommand{\mr}[1]{\mathrm{#1}}
\newcommand{\lf}{\left}
\newcommand{\ri}{\right}
\newcommand{\ds}{\displaystyle}
\newcommand{\dd}[2]{\frac{\rmd#1}{\rmd#2}}
\newcommand{\pp}[2]{\frac{\p #1}{\p #2}}

\newcommand{\de}[2]{\frac{\delta #1}{\delta #2}}
\newcommand{\delf}[2]{\frac{\delta #1\hfill}{\delta #2\hfill}}
\newcommand{\Div}{\nabla\cdot}
\newcommand{\Curl}{\nabla\times}
\newcommand{\nbl}{\nabla}
\makeatletter
\newcommand{\vast}{\bBigg@{4}}
\newcommand{\Vast}{\bBigg@{5}}
\makeatother

\newcommand{\calE}{\mathcal{E}}
\newcommand{\calF}{\mathcal{F}}

\newcommand{\calJ}{\mathcal{J}}

\newcommand{\scrF}{\mathscr{F}}
\newcommand{\scrG}{\mathscr{G}}
\newcommand{\scrH}{\mathscr{H}}

\DeclareMathAlphabet{\mathpzc}{OT1}{pzc}{m}{it}
\newcommand{\frakm}{\mathfrak{m}}

\newcommand{\rmc}{\mathrm{c}}
\newcommand{\rmd}{\mathrm{d}}
\newcommand{\rme}{\mathrm{e}}

\newcommand{\rmi}{\mathrm{i}}

\begin{document}

\title{Action Principles for Relativistic Extended Magnetohydrodynamics: A Unified Theory of Magnetofluid Models}

\author{Yohei Kawazura}
\email{yohei.kawazura@physics.ox.ac.uk}
\affiliation{Graduate School of Frontier Sciences, The University of Tokyo, Kashiwa, Chiba 277-8561, Japan}
\affiliation{Rudolf Peierls Centre for Theoretical Physics, University of Oxford, Oxford OX1 3NP, UK}

\author{George Miloshevich}
\email{gmilosh@physics.utexas.edu}
\affiliation{Department of Physics and Institute for Fusion Studies, The University of Texas at Austin, Austin, TX 78712, USA}

\author{Philip J. Morrison}
\email{morrison@physics.utexas.edu}
\affiliation{Department of Physics and Institute for Fusion Studies, The University of Texas at Austin, Austin, TX 78712, USA}

\date{\today}

% PACS, the Physics and Astronomy Classification Scheme.
% 95. Fundamental astronomy and astrophysics; instrumentation, techniques, and astronomical observations
	% 95.30.Qd	Magnetohydrodynamics and plasmas
	% 95.30.Sf	Relativity and gravitation

% 03. Quantum mechanics, field theories, and special relativity
	% 03.30.+p	Special relativity
	% 03.50.-z	Classical field theories

% 47. Fluid dynamics
	% 47.10.A-	Mathematical formulations
	% 47.75.+f	Relativistic fluid dynamics

% 52. Physics of plasmas and electric discharges
	% 52.27.Ny	Relativistic plasmas
	% 52.30.Cv	Magnetohydrodynamics

\begin{abstract}
Two types of Eulerian action principles for  relativistic extended magnetohydrodynamics (MHD) are formulated.  With the first,  the action is extremized under the constraints of density, entropy, and Lagrangian label conservation, which    leads to a Clebsch representation for a generalized momentum and a generalized vector potential.  
The second action arises upon transformation  to physical field variables,  giving rise to a covariant bracket action principle,  i.e., a variational principle  in which constrained variations are generated by a degenerate Poisson bracket. Upon taking appropriate limits, the action principles lead to relativistic Hall MHD and well-known relativistic ideal MHD.  
For the first time, the Hamiltonian formulation of relativistic Hall MHD  with 
electron thermal inertia (akin to [Comisso \textit{et al.}, Phys. Rev. Lett. {\bf 113}, 045001 (2014)] for  the electron--positron plasma) is introduced.  This  thermal inertia effect allows for violation of the frozen-in magnetic flux condition  in marked contrast to nonrelativistic Hall MHD that does satisfy the frozen-in condition.
We also find  violation of the frozen-in condition is accompanied by freezing-in of an  alternative flux determined by a generalized vector potential.
Finally,  we derive a more general 3+1 Poisson bracket for nonrelativistic extended MHD, one  that does not assume smallness of the electron  ion mass ratio.
\end{abstract}

\maketitle

%-------------------------------------------------------
% Introduction
%-------------------------------------------------------
\section{Introduction}
The early discovery of  action principles (AP)s and associated Hamiltonian structure, undoubtably of  groundbreaking importance in the history of physics,  has  unified existing physical models and provided a means for the development of new models. In physics it is now  believed that an empirically derived physical model, devoid of phenomenological constitutive relations,  would not be justified unless an underlying  AP exists.   In addition to mathematical elegance, APs are of practical importance for seeking  invariants  via  symmetries  using Noether's theorem~\cite{noether1918} (see, e.g.,  Refs.~\onlinecite{pjmP85,CB09} for plasma examples), obtaining consistent approximations (e.g.,   Ref.~\onlinecite{KeramidasCharidakos2014}), and developing numerical algorithms (e.g., Refs.~\onlinecite{Evstatiev:2013,pjmXQLYZH16, pjmKKS16}).

In this paper, we obtain  APs for relativistic magnetofluid models. The key  ingredient for constructing APs for a fluid-like  systems  is a means for implementing  constraints, because direct extremization yields trivial equations of motion. 
There are various  formalisms available,  depending on how the constraints are implemented.  One is to follow Lagrange~\cite{Lagrange1788} and  incorporate constraints into  the definition of the variables.  This procedure is invoked when using  Lagrangian coordinates  with the time evolution of variables (fluid element attributes) (e.g.,  density and entropy)  described a priori by conservation of differential forms along stream lines. APs in the Lagrangian coordinates have been obtained for the nonrelativistic neutral fluid, magnetohydrodynamics (MHD),\cite{Newcomb1962} and various generalized magnetofluid  models (e.g. extended MHD (XMHD), inertial MHD (IMHD), and Hall MHD (HMHD)),\cite{KeramidasCharidakos2014,Lingam2015a,DAvignon2016}  as well as for the relativistic neutral fluid~\cite{Taub1954,Dewar1977,Salmon1988} and MHD.\cite{Achterberg1983,Kawazura2014}  In obtaining such formulations  several complications arise, e.g.,  the inference of the appropriate Lagrangian variables, the map between the Lagrangian and Eulerian coordinates in the relativistic case,\cite{Kawazura2014} and the existence of multiple flow characteristics for generalized magnetofluid  models.\cite{KeramidasCharidakos2014,DAvignon2016} 

A second type of AP, one that  is formulated in terms Eulerian variables, implements the  constraints via  Lagrange multipliers, and in this way extremization  of the action can lead to  correct equations of motion.\cite{Serrin1959,Lin1963}  %, which we call the \emph{least constrained AP} hereafter.  
Upon enforcing the constraints  of conservation  of density, entropy, and  a Lagrangian label,\cite{Lin1963}  this procedure was recently used to obtain nonrelativistic HMHD.\cite{Yoshida2013}  Then, this  formulation for  HMHD  was  used to regularize  the  singular  limit to MHD  by a renormalization of variables,  thereby obtaining an  AP for MHD.\cite{Yoshida2013}   For the relativistic neutral fluid,  the velocity norm (lightcone) condition ($u^\mu u_\mu = 1$ with fluid four-velocity $u^\mu$) is required as another constraint.\cite{Schutz1970,Ray1972,Schutz1977,Kentwell1985,Elze1999}     Instead of taking the limit from HMHD with renormalization, there are alternative formulations for nonrelativistic~\cite{Webb2014} and relativistic~\cite{Bekenstein2000,Bekenstein2001} MHD, in which the Ohm's law or the induction equation \textit{per se} is employed as a constraint. 

A third type of AP, one  of general utility that  incorporates a covariant Poisson bracket in terms of  Eulerian variables,  was introduced in Ref.~\onlinecite{Marsden1986}. %, which we call the \emph{bracket AP} below.
Instead of including the constraints in the action with Lagrange multipliers, the constraints are implemented via the  degeneracy of a  Poisson bracket that effects constrained variations.   In addition to the neutral fluid, such Poisson bracket APs have been described for particle mechanics, electromagnetism, the Vlasov-Maxwell system, and the gravitational field.\cite{Marsden1986}      Most recently, this kind of action was obtained for   relativistic MHD.\cite{DAvignon2015} 

%-------------------------------------------------------
\begin{table*}[htpb]
\caption{Summary of APs for fluid-dynamical systems. The bold faces indicate APs which have not formulated until the present study.}
\label{t:knowns and unknowns}
	\centering
	{\renewcommand{\arraystretch}{1.2}%
	{\setlength{\tabcolsep}{0.3em}
	\begin{tabular}{|c|c|c|c|}
		\hline
		& Constrained least AP & Covariant bracket AP & Lagrangian description AP \\
		\hline
    Nonrelativistic    & \multirow{2}{*}{Lin (1963)}                                    & \multirow{2}{*}{\textbf{Present study}}           & \multirow{2}{*}{Lagrange (1788)}\\
		fluid               &                                                                &                                                   &                                 \\
		\hline
		Nonrelativistic    & Yoshida \& Hameiri (2013) (renormalization \& limit from HMHD) & \multirow{2}{*}{\textbf{Present study}}           & \multirow{2}{*}{Newcomb (1962)} \\ 
		MHD 	 		  		  	& Webb et al. (2014) (Ohm's law constraint)             &                                                   &                                 \\
		\hline
		Nonrelativistic    & \multirow{2}{*}{Yoshida \& Hameiri (2013) (HMHD)}              & \multirow{2}{*}{\textbf{Present study}}           & Keramidas Charidakos \\
		XMHD						  	&                                                                &                                                   & \hspace{4em} et al. (2014) \\
		\hline
		Relativistic        & \multirow{2}{*}{Schutz (1970)}                                  & \multirow{2}{*}{Marsden et al. (1986)}   & Dewar (1977)\\
		fluid						  	&                                                                &                                                   & Salmon (1988)\\
		\hline
		Relativistic        & \textbf{Present study (renormalization \& limit from HMHD)}    & \multirow{2}{*}{D'Avignon et al. (2015)} & Achterberg (1983)\\
		MHD								  & Bekenstein \& Oron (2000) (Ohm's law constraint)               &                                                   & Kawazura et al. (2014)\\
		\hline
		Relativistic        & \multirow{2}{*}{\textbf{Present study}}                        & \multirow{2}{*}{\textbf{Present study}}           & \multirow{2}{*}{\textbf{unknown}}\\
		XMHD							  &                                                                &                                                   &                         \\
		\hline
	\end{tabular}
	}
	}
\end{table*}
%-------------------------------------------------------

From Table~\ref{t:knowns and unknowns},  which summarizes the aforementioned APs, we see there are missing pieces:   the APs for fluid-dynamical systems are (i) the Lagrangian AP, (ii) Eulerian constrained least AP, and (iii) the Eulerian bracket AP, for relativistic generalized magnetofluid models.  In this paper, we formulate the latter two APs: (ii) and (iii), and show that they are related by variable transformation.
% Furthermore, reductions of the action derive relativistic HMHD and MHD. 
Then we derive APs for HMHD and MHD by taking limits of the  XMHD AP.  Relativistic HMHD is derived for the first time in the present study by this method. 
Also, we  show that the nonrelativistic limit of the bracket AP gives  nonrelativistic XMHD as a Hamiltonian system.

This paper is organized as follows.
In Sec.~\ref{s:constrained least AP} we formulate a constrained least AP for relativistic XMHD.
In Sec.~\ref{s:bracket AP} the bracket AP is derived by a transformation of phase space variables in the constrained least AP.
In Sec.~\ref{s:reduced models} we derive relativistic HMHD and MHD by taking  limits of  the bracket AP for XMHD. These results are used in Sec.~\ref{s:ColRecon}  where  remarkable features of  relativistic HMHD pertaining to collisionless reconnection are considered.
In Sec.~\ref{s:NR limit}, the nonrelativistic limit of the bracket AP is shown.
Finally in Sec.~\ref{s:conclusion} we conclude.  

%-------------------------------------------------------
% Constrained least action principle
%-------------------------------------------------------
\section{Constrained least action principle}
\label{s:constrained least AP}
Consider a relativistic plasma consisting of positively and negatively charged particles with  masses $m_+$ and $m_-$, where  subscript signs denote  species labels,  and assume  the  Minkowski spacetime with the metric tensor $\text{diag}(1,\, -1,\, -1,\, -1)$. 
In addition, a  proper charge neutrality condition is imposed so that rest frame particle number densities of each species satisfy $n_+ = n_- = n$.\cite{Koide2009}  The four-velocities of each species are denoted by ${u_\pm}^\mu$,  which obey the velocity norm conditions
%-------------------------------------------------------
\begin{eqnarray}
	{u_\pm}^\mu{u_\pm}_\mu = 1\,.
\label{e:velocity norm constraint}
\end{eqnarray}
%-------------------------------------------------------
Using the four-velocities ${u_\pm}^\mu$, the four-center of mass velocity and the four-current density can be  written as 
\begin{eqnarray}
u^\mu &=& (m_+/m) {u_+}^\mu + (m_-/m) {u_-}^\mu\,,
\label{e:umu}\\
 J^\mu &=&e({u_+}^\mu - {u_-}^\mu)\,,
 \label{e:jmu}
 \end{eqnarray}
 respectively,  with $m = m_+ + m_-$ and the electric charge $e$.
The time and space components of these fields are written as $u^\mu = (\gamma,\, \gamma\bm{v}/c)$ and $J^\mu = (\rho_q,\, \bm{J})$ with speed of light $c$, Lorentz factor $\gamma = 1/\sqrt{1 - (|\bm{v}|/c)^2}$, and charge density $\rho_q$.
The thermodynamic variables needed  are the energy density $\rho_\pm$, the enthalpy density $h_\pm$, the entropy density $\sigma_\pm$, and the isotropic pressure $p_\pm$. 
These are related by $nh_\pm = p_\pm + \rho_\pm = n(\p\rho_\pm/\p n) + \sigma_\pm(\p\rho_\pm/\p\sigma_\pm)$.\cite{Marsden1986} 
We also define total energy density $\rho = \rho_+ + \rho_-$ and total pressure $p = p_+ + p_-$.

Adding the continuity equations for each species together leads an equation for $n$, 
%-------------------------------------------------------
\begin{equation}
	\p_\nu(n u^\nu) = 0\,, 
\label{e:XMHD continuity}
\end{equation}
%-------------------------------------------------------
while the adiabatic equations of each species can be  written as 
%-------------------------------------------------------
\begin{eqnarray}
	\p_\nu\lf(\sigma_\pm {u_\pm}^\nu \ri) = 0.
\label{e:XMHD adiabaticity}
\end{eqnarray}
%-------------------------------------------------------
In addition to the above constraint equations we include  conservations of the Lagrangian labels $\varphi_\pm$, 
%-------------------------------------------------------
\begin{equation}
	{u_\pm}_\nu\p^\nu\varphi_\pm = 0.
\label{e:label conservation}
\end{equation}
%-------------------------------------------------------
The full set of  independent variables of our action are chosen to be $(u^\mu,\, J^\mu,\, n,\, \sigma_\pm,\, \varphi_\pm,\, A^\mu)$,  where $A^\mu$ is a four-vector potential that defines a Faraday tensor $\calF^{\mu\nu} = \p^\mu A^\nu - \p^\nu A^\mu$.
Here we consider CGS unit getting rid of a factor $1/4\pi$ in the Faraday tensor by renormalization (i.e., $\calF^{\mu\nu}/4\pi \to \calF^{\mu\nu}$).
In a manner similar to that of  Lin's formalism~\cite{Lin1963} for the nonrelativistic neutral fluid, we bring (\ref{e:XMHD continuity}), (\ref{e:XMHD adiabaticity}), and (\ref{e:label conservation}) into  an action as constraints as follows:
%-------------------------------------------------------
\begin{eqnarray}
	&& \hspace{-2em} S[u,\, J,\, n,\, \sigma_\pm,\, A,\, \varphi_\pm] = \nonumber\\
	&& \hspace{0.0em} \int\lf\{ \sum_{\pm}\lf[ -\f{1}{2}n h_\pm {u_\pm}_\nu{u_\pm}^\nu + \f{1}{2}(p_\pm - \rho_\pm) \ri] - J^\nu A_\nu \ri. \nonumber\\
	&& \hspace{1.5em} - \f{1}{4}(\p^\mu A^\nu - \p^\nu A^\mu)(\p_\mu A_\nu - \p_\nu A_\mu) - \phi\p^\nu(nu_\nu) \nonumber\\
	&& \hspace{3.5em} \lf.- \sum_{\pm}\lf[ \eta_\pm\p^\nu(\sigma_\pm {u_\pm}_\nu) - \lambda_\pm {u_\pm}_\nu\p^\nu\varphi_\pm \ri] \ri\}\rmd^4x,
	\label{e:XMHD S constrained}
\end{eqnarray}
where $\sum_{\pm}$ is summation over species, and $\phi$, $\eta_\pm$, and $\lambda_\pm$ are Lagrange multipliers. 
The first and second terms of \eqref{e:XMHD S constrained} are the fluid part for each species, the third term is an interaction between the fluid and the electromagnetic (EM) field, the fourth term is the pure EM part, and the other terms represent the constraints.
The velocity norm conditions (\ref{e:velocity norm constraint}) will be imposed after variation of the 
action.\cite{Note_on_velocity_norm_condition}

Variation of the action, i.e.,  setting $\delta S = 0$,  gives 
%-------------------------------------------------------
\begin{eqnarray}
	\label{e:XMHD delta u}
	\delta u_\nu &:& \; nhu^\nu + \f{\Delta h}{e}J^\nu = n\p^\nu\phi + \sum_{\pm}\lf( \sigma_\pm\p^\nu\eta_\pm + \lambda_\pm\p^\nu\varphi_\pm \ri) \\
	\label{e:XMHD delta J}
	\delta J_\nu &:& \; A^\nu + \f{\Delta h}{e}u^\nu + \f{h^\dagger}{ne^2}J^\nu = \nonumber\\
	& & \hspace{5em} \sum_{\pm}\lf[ \pm\f{m_\mp}{men}(\sigma_\pm\p^\nu\eta_\pm + \lambda_\pm\p^\nu\varphi_\pm) \ri] \\
	\label{e:XMHD delta sigma_+-}
	\delta \sigma_\pm&:& \; {u_\pm}_\nu\p^\nu\eta_\pm = \pp{\rho_\pm}{\sigma_\pm} \\
	\label{e:XMHD delta varphi_+-}
	\delta \varphi_\pm&:& \; \p^\nu\lf(\lambda_\pm {u_\pm}_\nu \ri) = 0\\
	\label{e:XMHD delta A}
	\delta A^\nu&:& \; J_\nu = \p^\mu \calF_{\mu\nu}\\
	\label{e:XMHD delta n}
	\delta n    &:& \; nu_\nu\p^\nu\phi = A^\nu J_\nu + n\sum_{\pm}\pp{\rho_\pm}{n},
\end{eqnarray}
%-------------------------------------------------------
with $h := h_+ + h_-$, $\Delta h := (m_-/m)h_+ - (m_+/m)h_-$, and $h^\dagger = (m_-^2/m^2)h_+ + (m_+^2/m^2)h_-$.
Using (\ref{e:XMHD continuity}), (\ref{e:XMHD adiabaticity}), (\ref{e:label conservation}), and (\ref{e:XMHD delta u})-(\ref{e:XMHD delta n}), the momentum equation and generalized Ohm's law are obtained: 
%-------------------------------------------------------
\begin{eqnarray}
	&& \p_\nu \lf[ nh u^\mu u^\nu + \f{\Delta h}{e}(u^\mu J^\nu + J^\mu u^\nu) + \f{h^\dagger}{ne^2}J^\mu J^\nu \ri]
\label{e:XMHD e.o.m.} 
	\\
	&&\hspace{.75cm}  = \p^\mu p + J^\nu {\calF^\mu}_{\nu}, 
	\nonumber\\
	&& \p_\nu \lf[ n(\Delta h) u^\mu u^\nu + \f{h^\dagger}{e}(u^\mu J^\nu + J^\mu u^\nu) + \f{\Delta h^\sharp}{ne^2}J^\mu J^\nu \ri] \label{e:XMHD Ohm's law} \\
	&& \hspace{.75cm}   = \f{m_-}{m}\p^\mu p_+ - \f{m_+}{m}\p^\mu p_- + enu^\nu {\calF^\mu}_{\nu} 
	- \f{m_+ - m_-}{m}J^\nu {\calF^\mu}_{\nu},
	\nonumber 
\end{eqnarray}
%-------------------------------------------------------
with $\Delta h^\sharp = (m_-^3/m^3)h_+ - (m_+^3/m^3)h_-$.
These are equivalent to the relativistic XMHD equations previously  formulated by Koide.\cite{Koide2009,Koide2010} 
The generalized Ohm's law of (\ref{e:XMHD e.o.m.}) can be  rewritten as
%-------------------------------------------------------
\begin{equation}
  eu_\nu{\calF^\star}^{\mu\nu} - \f{J_\nu}{n}{F^\dagger}^{\mu\nu} = \f{m_-}{m}\lf( T_+\p^\mu\f{\sigma_+}{n} \ri) - \f{m_+}{m}\lf( T_-\p^\mu\f{\sigma_-}{n} \ri),
\label{e:XMHD Ohm's law type2}
\end{equation}
%-------------------------------------------------------
with 
%-------------------------------------------------------
\begin{eqnarray*}
	{A^\dagger}^\nu &=& \f{m_+ - m_-}{m}A^\nu - \f{h^\dagger}{e}u^\nu - \f{\Delta h^\sharp}{ne^2}J^\nu, \\
	{\calF^\star}^{\mu\nu} &=& \p^\mu {A^\star}^\nu - \p^\nu {A^\star}^\mu \quad \text{and} \quad {\calF^\dagger}^{\mu\nu} = \p^\mu {A^\dagger}^\nu - \p^\nu {A^\dagger}^\mu\,, 
\end{eqnarray*}
%-------------------------------------------------------
where a generalized vector potential  $A^\star$ is defined by
\begin{equation}
  {A^\star}^\nu = A^\nu + \f{\Delta h}{e}u^\nu + \f{h^\dagger}{ne^2}J^\nu\,.
\label{e:starA}
\end{equation}
Note, the following must hold as an identity, 
%-------------------------------------------------------
\begin{equation}
	\p^\mu\lf( \epsilon_{\mu\nu\rho\sigma}{\calF^\star}^{\rho\sigma} \ri) = 0,
\label{e:dF* = 0}
\end{equation}
%-------------------------------------------------------
where $\epsilon_{\mu\nu\rho\sigma}$ is the  four-dimensional Levi-Civita symbol.
Upon taking the four-dimensional curl of (\ref{e:XMHD Ohm's law type2}), we obtain  the generalized induction equation
%-------------------------------------------------------
\begin{eqnarray}
	&& e\lf[ \p^\mu\lf( u_\lambda {\calF^\star}^{\nu\lambda} \ri) - \p^\nu\lf( u_\lambda {\calF^\star}^{\mu\lambda} \ri) \ri] - \lf[ \p^\mu\lf( \f{J_\lambda}{n}{\calF^\dagger}^{\nu\lambda} \ri) \ri. \nonumber\\ 
  && \lf. - \p^\nu\lf( \f{J_\lambda}{n}{\calF^\dagger}^{\mu\lambda} \ri) \ri] - \f{m_-}{m}\lf[ \p^\mu T_+\p^\nu\lf( \f{\sigma_+}{n} \ri) - \p^\nu T_+\p^\mu\lf( \f{\sigma_+}{n} \ri) \ri] \nonumber\\
	&& + \f{m_+}{m}\lf[ \p^\mu T_-\p^\nu\lf( \f{\sigma_-}{n} \ri) - \p^\nu T_-\p^\mu\lf( \f{\sigma_-}{n} \ri) \ri] = 0.
\label{e:XMHD induction eq.}
\end{eqnarray}
%-------------------------------------------------------

Next, upon combining  (\ref{e:XMHD e.o.m.}) and (\ref{e:XMHD Ohm's law}) we obtain  equations for the  canonical momenta~\cite{Mahajan2003} of  each species
%-------------------------------------------------------
\begin{eqnarray*}
	%{u_+}_\nu\lf( \p^\mu{\wp_+}^\nu - \p^\nu{\wp_+}^\mu \ri) + T_+\p^\mu\lf( \f{\sigma_+}{n} \ri) = 0,\\
	{u_{\pm}}_\nu\lf( \p^\mu{\wp_{\pm}}^\nu - \p^\nu{\wp_{\pm}}^\mu \ri) + T_{\pm}\p^\mu\lf( \f{\sigma_{\pm}}{n} \ri) = 0,
\end{eqnarray*}
%-------------------------------------------------------
where  ${\wp_{\pm}}^\nu = h_{\pm} {u_{\pm}}^\nu {\pm} eA^\nu$.
Several simplifications have been  proposed to make these equations tractable;\cite{Koide2009,Koide2010}  e.g., 
the  assumption of $\Delta h = 0$ (i.e.,  $h_+ = (m_+/m)h$ and $h_- = (m_-/m)h$) and/or the  usage of the velocity norm condition $u_\mu u^\mu = 1$ with \eqref{e:umu} instead of (\ref{e:velocity norm constraint}).   
The latter condition requires $J_\mu J^\mu = 0$ to be consistent with (\ref{e:velocity norm constraint}) (referred to as the  ``break down condition'' in Ref.~\onlinecite{Koide2009}).  Such a  simplified model has recently come into 
usage.\cite{Comisso2014,Asenjo2015,Asenjo2015a} 
Imposing $\Delta h = 0$ on  the action (\ref{e:XMHD S constrained}) and/or replacing (\ref{e:velocity norm constraint}) by  $u_\mu u^\mu = 1$ and $J_\mu J^\mu = 0$, this  simplified model is directly obtained  from the AP.

%-------------------------------------------------------
% Covariant bracket action principle
%-------------------------------------------------------
\section{Covariant bracket action principle}\label{s:bracket AP}
Now we construct our covariant action principle.  To this end we define  a kinetic momentum $\frakm^\nu = nhu^\nu$ and  a generalized momentum ${\frakm^\star}^\nu = \frakm^\nu + (\Delta h/e)J^\nu$.  Then (\ref{e:XMHD delta u}) and (\ref{e:XMHD delta J}) can then  be viewed   as the Clebsch representations for  ${\frakm^\star}^\nu$ and ${A^\star}^\nu$.  The reason for introducing these  new field variables  is that the action (\ref{e:XMHD S constrained}) takes a  beautiful form in terms of them: 
%-------------------------------------------------------
\begin{eqnarray}
  && S = \int \lf[ \f{{\frakm^\star}^\nu \frakm_\nu}{2nh} + \sum_\pm\f{1}{2}(p_\pm - \rho_\pm) \ri. 
		\label{e:XMHD S}\\
	&& \hspace{5em} \lf. - \f{1}{4}(\p^\mu {A^\star}^\nu - \p^\nu {A^\star}^\mu)(\p_\mu A_\nu - \p_\nu A_\mu) \ri]\rmd^4 x.
\nonumber
\end{eqnarray}
%-------------------------------------------------------
Interestingly, upon letting  ${\frakm^\star}^\nu \to \frakm^\nu$ and ${A^\star}^\nu \to A^\nu$, the action (\ref{e:XMHD S}) becomes identical to the recently  proposed relativistic MHD action of Ref.~\onlinecite{DAvignon2015}.
When the simplification $\Delta h \to 0$ is imposed, ${\frakm^\star}^\nu$ becomes the  kinetic momentum and ${A^\star}^\nu$ is decoupled from the kinetic momentum (the nonrelativistic version of such a vector potential was previously proposed for nonrelativistic IMHD~\cite{Lingam2015a} and XMHD~\cite{Abdelhamid2015,Lingam2015}).
In other words, the difference of the thermal inertiae between species (i.e. $\Delta h$) intertwines the kinetic momentum field and the  EM field.  Since the nonrelativistic limit ($h_+ \to m_+ c^2$ and $h_- \to m_- c^2$) results in $\Delta h \to 0$, such a coupling is distinctive of  the relativistic two-fluid plasma.

For our covariant action it is convenient to use  the Clebsch variables
\[
z=(n,\, \phi,\, \sigma_\pm,\, \eta_\pm,\, \lambda_\pm,\, \varphi_\pm)
\]
as   the independent variables of the action (\ref{e:XMHD S}).
With these variables all of the dynamical equations (\ref{e:XMHD continuity}), (\ref{e:XMHD adiabaticity}), (\ref{e:label conservation}), (\ref{e:XMHD delta sigma_+-}), (\ref{e:XMHD delta varphi_+-}), and (\ref{e:XMHD delta n}) are derived from the least AP (i.e. $\delta S = 0$).   In terms of $z$ we can simply restate the  AP of Sec.~\ref{s:constrained least AP} as a canonical  covariant bracket version of the formalism of Refs.~\onlinecite{Marsden1986,DAvignon2015}.
A canonical Poisson bracket is defined for functionals $F$ and $G$ as
%-------------------------------------------------------
\begin{eqnarray}
&&\lf\{ F,\, G \ri\}_\mathrm{canonical} = \int \de{F}{z} \calJ_\rmc \de{G}{z}\,  \rmd^4x
\nonumber\\
&&\hspace{1.5 cm}= \int\Bigg[ \delf{F}{\phi}\delf{G}{n} - \delf{G}{\phi}\delf{F}{n} 
\label{e:XMHD canonical bracket}\\
&&\hspace{2.5 cm} + \sum_{\pm}\lf( \delf{F}{\eta_\pm}\delf{G}{\sigma_\pm} - \delf{G}{\eta_\pm}\delf{F}{\sigma_\pm}
\right. \nonumber\\
&&\hspace{3.5 cm}+ \left. \delf{F}{\varphi_\pm}\delf{G}{\lambda_\pm} - \delf{G}{\varphi_\pm}\delf{F}{\lambda_\pm} \ri) \Bigg] \, \rmd^4x,
\nonumber
\end{eqnarray}
%-------------------------------------------------------
where $\calJ_\rmc$ is the symplectic matrix and $\delta F/\delta z$ denotes the functional derivative obtained by linearizing a functional,  e.g.\  
\begin{equation}
\delta F= \int  \delta n \, \de{F}{n} \,  \rmd^4x\,.
\label{e:4FD}
\end{equation}
(See Ref.~\onlinecite{pjm98} for review.) Since $\calJ_\rmc$ is non-degenerate, with $(n,\phi)$, $(\sigma_\pm,\eta_\pm)$, and $(\lambda_\pm,\varphi_\pm)$ being canonically conjugate pairs, the least AP is equivalent to a bracket AP, i.e.,  $\{ F[z], \, S \}_\mathrm{canonical} = 0$  where $F[z]$ is an arbitrary functional of $z$,  is equivalent to $\delta S=0$.

Transformation  to new ``physical''  independent variables  defined by 
\[
\bar{z} = (n, \sigma_\pm, {\frakm^\star}^\nu, {\calF^\star}^{\mu\nu})\]
yields a noncanonical covariant bracket because the $\bar{z}$ are not canonical variables. 
To transform the bracket of  \eqref{e:XMHD canonical bracket} we consider  functionals that satisfy $\bar{F}[\bar{z}] = F[z]$, and calculate functional derivatives  by the  chain rule
%-------------------------------------------------------
\begin{eqnarray*}
	\delf{F}{n} &=& \delf{\bar{F}}{n} + \delf{\bar{F}}{{\frakm^\star}^\nu}\p^\nu\phi\\
	& & + \f{2m_-\sigma_+}{men^2}\p^\mu\lf( \delf{\bar{F}}{{\calF^\star}^{\mu\nu}}\p^\nu\eta_+ \ri) + \f{2m_-\lambda_+}{men^2}\p^\mu\lf( \delf{\bar{F}}{{\calF^\star}^{\mu\nu}}\p^\nu\varphi_+ \ri) \\
	& & - \f{2m_+\sigma_-}{men^2}\p^\mu\lf( \delf{\bar{F}}{{\calF^\star}^{\mu\nu}}\p^\nu\eta_- \ri) - \f{2m_+\lambda_-}{men^2}\p^\mu\lf( \delf{\bar{F}}{{\calF^\star}^{\mu\nu}}\p^\nu\varphi_- \ri), \\
	\\
	\delf{F}{\phi} &=& -\p^\nu\lf( n\delf{\bar{F}}{{\frakm^\star}^\nu} \ri), \\
	\\
	\delf{F}{\sigma_\pm} &=& \delf{\bar{F}}{\sigma_\pm} + \delf{\bar{F}}{{\frakm^\star}^\nu}\p^\nu\eta_\pm \mp \f{2m_\mp}{men}\p^\mu\lf( \delf{\bar{F}}{{\calF^\star}^{\mu\nu}}\p^\nu\eta_\pm \ri), \\
	\\
	\delf{F}{\eta_\pm} &=& -\p^\nu\lf( \sigma_\pm\delf{\bar{F}}{{\frakm^\star}^\nu} \ri) \pm \f{2m_\mp}{me}\p^\mu\lf( \delf{\bar{F}}{{\calF^\star}^{\mu\nu}}\p^\nu\f{\sigma_\pm}{n} \ri), \\
	\\
	\delf{F}{\lambda_\pm} &=& \delf{\bar{F}}{{\frakm^\star}^\nu}\p^\nu\varphi_\pm \mp \f{2m_\mp}{men}\p^\mu\lf( \delf{\bar{F}}{{\calF^\star}^{\mu\nu}}\p^\nu\varphi_\pm \ri), \\
	\\
	\delf{F}{\varphi_\pm} &=& -\p^\nu\lf( \lambda_\pm\delf{\bar{F}}{{\frakm^\star}^\nu} \ri) \pm \f{2m_\mp}{me}\p^\mu\lf( \delf{\bar{F}}{{\calF^\star}^{\mu\nu}}\p^\nu\f{\lambda_\pm}{n} \ri) \\
\end{eqnarray*}
%-------------------------------------------------------
Substituting these into (\ref{e:XMHD canonical bracket}) gives the following  noncanonical Poisson bracket, 
%-------------------------------------------------------
\begin{eqnarray}
	\lf\{ \bar{F},\, \bar{G} \ri\}_\mathrm{XMHD} &=& -\int\lf\{ 
	n\lf( \delf{\bar{G}}{{\frakm^\star}^\nu}\p^\nu\delf{\bar{F}}{n} - \delf{\bar{F}}{{\frakm^\star}^\nu}\p^\nu\delf{\bar{G}}{n} \ri) \right.
	\nonumber\\ 
 &+&  \left. {\frakm^\star}^\nu\lf( \delf{\bar{G}}{{\frakm^\star}^\mu}\p^\mu\delf{\bar{F}}{{\frakm^\star}^\nu} 
	- \delf{\bar{F}}{{\frakm^\star}^\mu}\p^\mu\delf{\bar{G}}{{\frakm^\star}^\nu} \ri) \ri. 
 \nonumber \\
  \nonumber \\
 &+&   \sum_{\pm} \lf[ \sigma_\pm 
 \lf( \delf{\bar{G}}{{\frakm^\star}^\nu}\p^\nu\delf{\bar{F}}{\sigma_\pm} 
 - \delf{\bar{F}}{{\frakm^\star}^\nu}\p^\nu\delf{\bar{G}}{\sigma_\pm} \ri) \right.
\nonumber\\ 
& &\pm\left.
 \f{2m_\mp}{me}\lf( \delf{\bar{F}}{\sigma_\pm}\p^\mu\delf{\bar{G}}{{\calF^\star}^{\mu\nu}} - \delf{\bar{G}}{\sigma_\pm}\p^\mu\delf{\bar{F}}{{\calF^\star}^{\mu\nu}} \ri)\p^\nu\f{\sigma_\pm}{n} \ri] \nonumber\\
	\nonumber \\
	&+&  \lf.  2\lf( \delf{\bar{F}}{{\frakm^\star}^\lambda}\p^\mu\delf{\bar{G}}{{\calF^\star}^{\mu\nu}} - \delf{\bar{G}}{{\frakm^\star}^\lambda}\p^\mu\delf{\bar{F}}{{\calF^\star}^{\mu\nu}} \ri){\calF^\star}^{\nu\lambda} \right.
	\nonumber \\
	& +&
	\left. \f{4}{ne}\lf( \p^\mu\delf{\bar{F}}{{\calF^\star}^{\mu\nu}} \ri)\lf( \p^\lambda\delf{\bar{G}}{{\calF^\star}^{\lambda\kappa}} \ri){\calF^\dagger}^{\kappa\nu} \ri\}\rmd^4x\,. 
\label{e:XMHD noncanonical bracket}
\end{eqnarray}
%-------------------------------------------------------
The fluid  parts (the first three terms) of \eqref{e:XMHD noncanonical bracket} correspond to the covariant Poisson bracket for the neutral  fluid given in  Ref.~\onlinecite{Marsden1986}. 
Next, in order to use this bracket  in a variational sense, the  action (\ref{e:XMHD S}) is considered to be the functional of ($n, \sigma_\pm, {\frakm^\star}^\nu, {\calF^\star}^{\mu\nu}$), i.e. $\bar{S}[\bar{z}]$,
and its functional derivatives are calculated as
%-------------------------------------------------------
\begin{eqnarray*}
	&& \delf{\bar{S}}{{\frakm^\star}^{\nu}} = u_\nu, \quad \delf{\bar{S}}{{\calF^\star}^{\mu\nu}} = -\f{1}{2}\calF_{\mu\nu}, \quad \delf{\bar{S}}{\sigma_\pm} = -\pp{\rho_\pm}{\sigma_\pm} \\
	\\
	&& \delf{\bar{S}}{n} = h_+\f{m_-}{men}J_\nu \lf( u^\nu + \f{m_-}{men}J^\nu \ri) - h_-\f{m_+}{men}J_\nu \lf( u^\nu - \f{m_+}{men}J^\nu \ri) \\
	&& \hspace{2.5em} -\pp{\rho_+}{n} - \pp{\rho_-}{n} + \f{\Delta h}{ne}J_\nu u^\nu + \f{h^\dagger}{n^2 e^2}J_\nu J^\nu.
\end{eqnarray*}
%-------------------------------------------------------
Then equations (\ref{e:XMHD continuity}), (\ref{e:XMHD adiabaticity}), (\ref{e:XMHD e.o.m.}), and (\ref{e:XMHD induction eq.})  follow from  $\{\bar{F}[\bar{z}],\, \bar{S} \} = 0$ for all $\bar{F}$.

Here we must remark that the equations obtained from the bracket action principle are not closed unless (\ref{e:dF* = 0}) is imposed.
Although (\ref{e:dF* = 0}) is automatically satisfied by the Clebsch variable definition of ${\calF^\star}^{\mu\nu}$, it does not emerge from the bracket AP. 
% Since (\ref{e:dF* = 0}) is dynamical, we may not solve the system without it.
Therefore, the bracket AP, only by itself, does not give the closed set of equations.
This is a marked difference between a Hamiltonian formalism of nonrelativistic MHD;\cite{Morrison1982a,Morrison1982b} 
although $\Div \bm{B} = 0$ is not derived from the Hamiltonian equation, the obtained equations are closed even if $\Div \bm{B} \ne 0$.
On the other hand, in the relativistic case, if (\ref{e:dF* = 0}) is abandoned, we lose $\p_t \bm{B} = -c\Curl\bm{E}$ as well.  

There may be two remedies for this problem.
One is to define a Faraday tensor  that builds-in the Ohm's law (\ref{e:XMHD Ohm's law type2}) and consider (\ref{e:dF* = 0}) as a dynamical equation of the new Faraday tensor~\cite{Anile1990, DAvignon2015}.
This strategy, however, is difficult because the Ohm's law (\ref{e:XMHD Ohm's law type2}) is more complicated than that of relativistic MHD, and then it is hard to formulate the appropriate Faraday tensor.
A second approach is to transform  ${\calF^\star}^{\mu\nu}$ to ${A^\star}^\mu$ so as to make the bracket action principle yield Ohm's law instead of the induction equation. To write the bracket of \eqref{e:XMHD noncanonical bracket} in terms of ${A^\star}^\mu$ we consider the functional chain rule to relate functional derivatives with respect to ${\calF^\star}^{\mu\nu}$  with those with respect to ${A^\star}^\mu$, i.e. 
%-------------------------------------------------------
\begin{equation}
	2\p_\nu\delf{\bar{G}}{{\calF^\star}^{\mu\nu}} = \delf{\bar{G}}{{A^\star}^{\mu}}.
	\label{e:AtoF}
\end{equation}
%-------------------------------------------------------
Using \eqref{e:AtoF},  one can eliminate ${\calF^\star}^{\mu\nu}$ from the Poisson bracket (\ref{e:XMHD noncanonical bracket}) while introducing the variable ${A^\star}^\mu$.  This will give a bracket where the Ohm's law (\ref{e:XMHD Ohm's law type2}) is obtained directly.  The transformation \eqref{e:AtoF} yields
%-------------------------------------------------------
% \begin{widetext}
% \begin{eqnarray}
	% \lf\{ \bar{F},\, \bar{G} \ri\}_\mathrm{XMHD} &=& \int\lf\{ n\lf( \delf{\bar{G}}{{\frakm^\star}^\nu}\p^\nu\delf{\bar{F}}{n} - \delf{\bar{F}}{{\frakm^\star}^\nu}\p^\nu\delf{\bar{G}}{n} \ri) + {\frakm^\star}^\nu\lf( \delf{\bar{G}}{{\frakm^\star}^\mu}\p^\mu\delf{\bar{F}}{{\frakm^\star}^\nu} - \delf{\bar{F}}{{\frakm^\star}^\mu}\p^\mu\delf{\bar{G}}{{\frakm^\star}^\nu} \ri) \ri. \nonumber\\
	% \nonumber \\
	% & & \hspace{2em} + \sum_{\pm} \lf[ \sigma_\pm\lf( \delf{\bar{G}}{{\frakm^\star}^\nu}\p^\nu\delf{\bar{F}}{\sigma_\pm} - \delf{\bar{F}}{{\frakm^\star}^\nu}\p^\nu\delf{\bar{G}}{\sigma_\pm} \ri) \mp \f{m_\mp}{me}\lf( \delf{\bar{F}}{\sigma_\pm}\delf{\bar{G}}{{A^\star}^{\mu}} - \delf{\bar{G}}{\sigma_\pm}\delf{\bar{F}}{{A^\star}^{\mu}} \ri)\p^\nu\f{\sigma_\pm}{n} \ri] \nonumber\\
	% \nonumber \\
	% & & \hspace{2em} \lf. + \lf( \delf{\bar{G}}{{\frakm^\star}^\nu}\delf{\bar{F}}{{A^\star}^{\mu}} - \delf{\bar{F}}{{\frakm^\star}^\nu}\delf{\bar{G}}{{A^\star}^{\mu}} \ri){\calF^\star}^{\mu\nu} - \f{1}{ne}\delf{\bar{F}}{{A^\star}^{\mu}}\delf{\bar{G}}{{A^\star}^{\nu}}{\calF^\dagger}^{\mu\nu} \ri\}\rmd^4x. 
% \label{e:XMHD noncanonical bracket with A}
% \end{eqnarray}
% \end{widetext}
% \begin{widetext}
\begin{eqnarray}
	\lf\{ \bar{F},\, \bar{G} \ri\}_\mathrm{XMHD} 
  &=& -\int\Bigg\{ n\lf( \delf{\bar{G}}{{\frakm^\star}^\nu}\p^\nu\delf{\bar{F}}{n} - \delf{\bar{F}}{{\frakm^\star}^\nu}\p^\nu\delf{\bar{G}}{n} \ri) \nonumber\\
  &+& {\frakm^\star}^\nu\lf( \delf{\bar{G}}{{\frakm^\star}^\mu}\p^\mu\delf{\bar{F}}{{\frakm^\star}^\nu} - \delf{\bar{F}}{{\frakm^\star}^\mu}\p^\mu\delf{\bar{G}}{{\frakm^\star}^\nu} \ri) \nonumber\\
  \nonumber \\
  &+& \sum_{\pm} \Bigg[ \sigma_\pm\lf( \delf{\bar{G}}{{\frakm^\star}^\nu}\p^\nu\delf{\bar{F}}{\sigma_\pm} - \delf{\bar{F}}{{\frakm^\star}^\nu}\p^\nu\delf{\bar{G}}{\sigma_\pm} \ri) \nonumber\\
  & &\mp \f{m_\mp}{me}\lf( \delf{\bar{F}}{\sigma_\pm}\delf{\bar{G}}{{A^\star}^{\nu}} - \delf{\bar{G}}{\sigma_\pm}\delf{\bar{F}}{{A^\star}^{\nu}} \ri)\p^\nu\f{\sigma_\pm}{n} \Bigg] \nonumber\\
	\nonumber \\
	&+& \lf( \delf{\bar{G}}{{\frakm^\star}^\nu}\delf{\bar{F}}{{A^\star}^{\mu}} - \delf{\bar{F}}{{\frakm^\star}^\nu}\delf{\bar{G}}{{A^\star}^{\mu}} \ri){\calF^\star}^{\mu\nu} \nonumber\\
  &-& \f{1}{ne}\delf{\bar{F}}{{A^\star}^{\mu}}\delf{\bar{G}}{{A^\star}^{\nu}}{\calF^\dagger}^{\mu\nu} \Bigg\}\rmd^4x. 
\label{e:XMHD noncanonical bracket with A}
\end{eqnarray}
%-------------------------------------------------------
Ohm's law  follows from  $\{ {A^\star}^\alpha,\, \bar{S} \}_\mathrm{XMHD} = 0$ with  $\delta \bar{S}/\delta{A^\star}^\mu = J_\mu$;  the  other equations are unaltered so  the system is  closed.

The  noncanonical bracket of (\ref{e:XMHD noncanonical bracket with A})  has the form 
%-------------------------------------------------------
\begin{eqnarray*}
	\lf\{ \bar{F},\, \bar{G} \ri\}_\mathrm{XMHD}=\int \de{\bar{F}}{\bar{z}}\, \calJ\, \de{\bar{G}}{\bar{z}} \rmd^4x\,,
\end{eqnarray*}
%-------------------------------------------------------
with a new Poisson operator $\calJ$.  However, because the transformation $z\mapsto \bar{z}$ is not invertible, the Poisson operator   $\cal J$ is degenerate.    Since the bracket AP,  $\{ \bar{F}[\bar{z}],\, \bar{S} \} = 0$,  is equivalent to $\calJ \delta \bar{S}/\delta \bar{z} = 0$, because of this degeneracy it is no longer true that $\calJ \,\delta \bar{S}/\delta \bar{z} = 0$ is  identical to  $\delta S = 0$.   In this way the  constraints of  the action (\ref{e:XMHD S constrained}) are transferred  to the degeneracy of the Poisson bracket.\cite{Marsden1986,DAvignon2015} 

Before closing this section, let us make a remark about the alternative expression of the EM field.
The Faraday tensor may be decomposed as $\calF^{\mu\nu} = \epsilon^{\mu\nu\lambda\sigma}b_\lambda u_\sigma + u^\mu e^\nu - u^\nu e^\mu$ with a magnetic field like four vector $b^\nu = u_\mu \epsilon^{\mu\nu\lambda\sigma}\calF_{\lambda\sigma}$ and a electric field like four vector $e^\nu = u_\mu \calF^{\mu\nu}$.~\cite{Lichnerowicz1967,Anile1990,Pegoraro2015}
This decomposition is especially useful in the relativistic MHD because the standard Ohm's law is equivalent to $e^\nu = 0$, and thus EM field is concisely expressed only by $b^\nu$. 
In the context of the action principle, D'Avignon \textit{et al.} formulated the bracket AP for the relativistic MHD using $b^\nu$.~\cite{DAvignon2015}
It may be possible to reformulate the relativistic XMHD action principle in terms of $b^\nu$ instead of $\calF^{\mu\nu}$.
The key is how we define a generalized four vector (say $b^{\star\nu}$) that incorporates inertia effect in the similar way as $\calF^{\mu\nu} \to \calF^{\star\mu\nu}$. 
Recently, such a generalization of $b^\nu$ has been proposed by Pegoraro.~\cite{Pegoraro2015}
Formulation of the action principle with $b^{\star\nu}$ and the unification with the MHD action principle~\cite{DAvignon2015} will be a future work.

%-------------------------------------------------------
% Limit to reduced models
%-------------------------------------------------------
\section{Limits to reduced models}
\label{s:reduced models}

In this section we show how to reduce the  bracket AP to obtain APs for unknown  relativistic  models, with known  nonrelativistic counterparts.      

First consider the electron-ion plasma, where now  the species labels $+$ and $-$ are replaced by i and e, respectively.
Defining electron to ion mass ratio $\mu := m_\rme/m_\rmi \ll 1$, we approximate $m_\rme/m \sim \mu$, $m_\rmi/m \sim 1$.
The ion and electron four-velocities become
%-------------------------------------------------------
\begin{eqnarray*}
	u_\rmi^\nu = u^\nu + \f{\mu J^\nu}{ne}, \quad u_\rme^\nu = u^\nu - \f{J^\nu}{ne}\,,
\end{eqnarray*}
%-------------------------------------------------------
while the enthalpy variables  reduce to $\Delta h \sim \mu h_\rmi - h_\rme$,  $h^\dagger \sim \mu^2 h_\rmi + h_\rme$, and $\Delta h^\sharp \sim \mu^3 h_\rmi - h_\rme$,   
%-------------------------------------------------------
and the generalized vectors become
%-------------------------------------------------------
\begin{eqnarray}
  \label{e:mstar electron-ion} 
	&&{\frakm^\star}^\nu = nhu^\nu + \f{1}{e}(\mu h_\rmi - h_\rme)J^\nu \\
  \label{e:Astar electron-ion} 
	&&{A^\star}^\nu = A^\nu + \f{1}{e}(\mu h_\rmi - h_\rme)u^\nu + \f{1}{ne^2}(\mu^2 h_\rmi + h_\rme)J^\nu \\
  \label{e:Adagger electron-ion} 
	&&{A^\dagger}^\nu = A^\nu - \f{1}{e}(\mu^2 h_\rmi + h_\rme)u^\nu - \f{1}{ne^2}(\mu^3 h_\rmi - h_\rme)J^\nu.
\end{eqnarray}
%-------------------------------------------------------
Next, in this approximation the noncanonical Poisson bracket (\ref{e:XMHD noncanonical bracket with A}) becomes
\begin{eqnarray}
	\lf\{ \bar{F},\, \bar{G} \ri\}_\mathrm{XMHD} 
  &=& -\int\Bigg\{ n\lf( \delf{\bar{G}}{{\frakm^\star}^\nu}\p^\nu\delf{\bar{F}}{n} - \delf{\bar{F}}{{\frakm^\star}^\nu}\p^\nu\delf{\bar{G}}{n} \ri) \nonumber\\
  &+& {\frakm^\star}^\nu\lf( \delf{\bar{G}}{{\frakm^\star}^\mu}\p^\mu\delf{\bar{F}}{{\frakm^\star}^\nu} - \delf{\bar{F}}{{\frakm^\star}^\mu}\p^\mu\delf{\bar{G}}{{\frakm^\star}^\nu} \ri) \nonumber\\
  \nonumber \\
	&+& \sigma_\rmi\lf( \delf{\bar{G}}{{\frakm^\star}^\nu}\p^\nu\delf{\bar{F}}{\sigma_\rmi} - \delf{\bar{F}}{{\frakm^\star}^\nu}\p^\nu\delf{\bar{G}}{\sigma_\rmi} \ri) \nonumber\\
  &-& \f{\mu}{e}\lf( \delf{\bar{F}}{\sigma_\rmi}\delf{\bar{G}}{{A^\star}^{\nu}} - \delf{\bar{G}}{\sigma_\rmi}\delf{\bar{F}}{{A^\star}^{\nu}} \ri)\p^\nu\lf( \f{\sigma_\rmi}{n} \ri) \nonumber\\
  \nonumber \\
	&+& \sigma_\rme\lf( \delf{\bar{G}}{{\frakm^\star}^\nu}\p^\nu\delf{\bar{F}}{\sigma_\rme} - \delf{\bar{F}}{{\frakm^\star}^\nu}\p^\nu\delf{\bar{G}}{\sigma_\rme} \ri) \nonumber\\
  &+& \f{1}{e}\lf( \delf{\bar{F}}{\sigma_\rme}\delf{\bar{G}}{{A^\star}^{\nu}} - \delf{\bar{G}}{\sigma_\rme}\delf{\bar{F}}{{A^\star}^{\nu}} \ri)\p^\nu\lf( \f{\sigma_\rme}{n} \ri) \nonumber\\
	\nonumber \\
	&+& \lf( \delf{\bar{G}}{{\frakm^\star}^\nu}\delf{\bar{F}}{{A^\star}^{\mu}} - \delf{\bar{F}}{{\frakm^\star}^\nu}\delf{\bar{G}}{{A^\star}^{\mu}} \ri){\calF^\star}^{\mu\nu} \nonumber\\
  &-& \f{1}{ne}\delf{\bar{F}}{{A^\star}^{\mu}}\delf{\bar{G}}{{A^\star}^{\nu}}{\calF^\dagger}^{\mu\nu} \Bigg\}\rmd^4x. 
\label{e:XMHD noncanonical bracket electron-ion with A}
\end{eqnarray}
%-------------------------------------------------------
Using the approximate  bracket  of \eqref{e:XMHD noncanonical bracket electron-ion with A} with a reduced action $\bar{S}$, the covariant   AP produces  the continuity equation (\ref{e:XMHD continuity}) along with  the following system of equations:
%-------------------------------------------------------
\begin{eqnarray}
	&& \p_\nu\lf[ nhu^\mu u^\nu + \f{1}{e}(\mu h_\rmi - h_\rme)(u^\mu J^\nu + J^\mu u^\nu) \ri. \nonumber\\
	&& \hspace{3em} \lf. + \f{1}{ne^2}(\mu^2 h_\rmi + h_\rme)J^\mu J^\nu \ri] = \p^\mu p + J_\nu {\calF^{\mu\nu}}, 
	\label{e:XMHD e.o.m. electron-ion} \\
	\nonumber\\
	&& eu_\nu{\calF^\star}^{\mu\nu} - \f{J_\nu}{n}{\calF^\dagger}^{\mu\nu} - \mu T_\rmi\p^\mu\lf( \f{\sigma_\rmi}{n} \ri) + T_\rme\p^\mu\lf( \f{\sigma_\rme}{n} \ri) = 0,
	\label{e:XMHD Ohm's law electron-ion} \\
	\nonumber\\
	&& \p_\nu\lf[\sigma_\rmi \lf( u^\nu + \f{\mu J^\nu}{ne} \ri) \ri] = 0,
	\label{e:XMHD adiabaticity i electron-ion} \\
	\nonumber\\
	&& \p_\nu\lf[\sigma_\rme \lf( u^\nu - \f{J^\nu}{ne} \ri)\ri] = 0.
	\label{e:XMHD adiabaticity e electron-ion} 
\end{eqnarray}
%-------------------------------------------------------

Next consider a  further reduction using  $\mu \to 0$, meaning the electron rest mass inertia is discarded.
This limit gives  HMHD,  which is well known in the nonrelativistic case but has not been proposed in the relativistic case.
The terms including $h_\rme$ must  not be ignored when the electron thermal inertia is greater than the rest mass inertia (i.e., $h_\rme \gg m_\rme c^2$). 
For example, the temperature of  electrons in an  accretion disk near a black hole can be more  than $10^{11}\,\mr{K}$.\cite{yuan2002jet} 
Then, the thermal inertia $h_\rme$ is on the order of $100\,m_\rme c^2$, estimated by an equation of state for an  ideal gas $h_\rme = m_\rme c^2 + [\Gamma/(\Gamma - 1)]T_\rme$ with a specific heat ratio $\Gamma = 4/3$.\cite{taub1948relativistic}
In such a case, the $h_\rme$ terms are not negligible.

Let us employ the  following normalizations: 
%-------------------------------------------------------
\begin{eqnarray*}
	\p^\nu \to L^{-1} \p^\nu, \quad n \to n_0 n, \quad T_{\rmi,\rme} \to mc^2 T_{\rmi,\rme}, \\
	\quad \sigma_{\rmi,\rme} \to n_0 \sigma_{\rmi,\rme}, \quad \calF^{\mu\nu} \to \sqrt{n_0 mc^2}\calF^{\mu\nu},
\end{eqnarray*}
%-------------------------------------------------------
using a  typical scale length $L$ and density scale $n_0$.
Then the  generalized momentum density and vector potential are normalized as
%-------------------------------------------------------
\begin{eqnarray*}
	&& {\frakm^\star}^\nu \to n_0 m c^2 \lf[ nhu^\nu - d_\rmi h_\rme J^\nu \ri]
	\label{e:normalized m} \\
	&& {A^\star}^\nu \to L\sqrt{n_0 m c^2}\lf[ A^\nu - d_\rmi h_\rme u^\nu + {d_\rmi}^2h_\rme\f{J^\nu}{n} \ri],
	\label{e:normalized A*} 
\end{eqnarray*}
%-------------------------------------------------------
where $\sqrt{(m c^2)/(e^2 n_0 L^2)} \sim \sqrt{(m_\rmi c^2)/(e^2 n_0 L^2)} = c/(\omega_\rmi L) = d_\rmi$ is the normalized ion skin depth, and the  normalized Poisson bracket becomes 
%-------------------------------------------------------
% \begin{widetext}
% \begin{eqnarray}
	% \lf\{ \bar{F},\, \bar{G} \ri\}_\mathrm{HMHD} &=& \int\lf\{ n\lf( \delf{\bar{G}}{{\frakm^\star}^\nu}\p^\nu\delf{\bar{F}}{n} - \delf{\bar{F}}{{\frakm^\star}^\nu}\p^\nu\delf{\bar{G}}{n} \ri) + {\frakm^\star}^\nu\lf( \delf{\bar{G}}{{\frakm^\star}^\mu}\p^\mu\delf{\bar{F}}{{\frakm^\star}^\nu} - \delf{\bar{F}}{{\frakm^\star}^\mu}\p^\mu\delf{\bar{G}}{{\frakm^\star}^\nu} \ri) \ri. \nonumber\\
	% \nonumber \\
	% & & \hspace{2em} + \sigma_\rmi\lf( \delf{\bar{G}}{{\frakm^\star}^\nu}\p^\nu\delf{\bar{F}}{\sigma_\rmi} - \delf{\bar{F}}{{\frakm^\star}^\nu}\p^\nu\delf{\bar{G}}{\sigma_\rmi} \ri) \nonumber\\
	% \nonumber \\
	% & & \hspace{2em} + \sigma_\rme\lf( \delf{\bar{G}}{{\frakm^\star}^\nu}\p^\nu\delf{\bar{F}}{\sigma_\rme} - \delf{\bar{F}}{{\frakm^\star}^\nu}\p^\nu\delf{\bar{G}}{\sigma_\rme} \ri) - 2d_\rmi\lf( \delf{\bar{F}}{\sigma_\rme}\p^\mu\delf{\bar{G}}{{\calF^\star}^{\mu\nu}} - \delf{\bar{G}}{\sigma_\rme}\p^\mu\delf{\bar{F}}{{F^\star}^{\mu\nu}} \ri)\p^\nu\f{\sigma_\rme}{n} \nonumber\\
	% \nonumber \\
	% & & \hspace{2em} \lf. + 2\lf( \delf{\bar{F}}{{\frakm^\star}^\lambda}\p^\mu\delf{\bar{G}}{{\calF^\star}^{\mu\nu}} - \delf{\bar{G}}{{\frakm^\star}^\lambda}\p^\mu\delf{\bar{F}}{{\calF^\star}^{\mu\nu}} \ri){\calF^\star}^{\nu\lambda} + \f{4d_\rmi}{n}\lf( \p^\mu\delf{\bar{F}}{{\calF^\star}^{\mu\nu}} \ri)\lf( \p^\lambda\delf{\bar{G}}{{\calF^\star}^{\lambda\kappa}} \ri){\calF^\star}^{\kappa\nu} \ri\}\rmd^4x. 
% \label{e:XMHD noncanonical bracket electron-ion}
% \end{eqnarray}
% \end{widetext}
% \begin{widetext}
\begin{eqnarray}
	\lf\{ \bar{F},\, \bar{G} \ri\}_\mathrm{HMHD} &=& -\int\Bigg\{ n\lf( \delf{\bar{G}}{{\frakm^\star}^\nu}\p^\nu\delf{\bar{F}}{n} - \delf{\bar{F}}{{\frakm^\star}^\nu}\p^\nu\delf{\bar{G}}{n} \ri) \nonumber\\
  &+& {\frakm^\star}^\nu\lf( \delf{\bar{G}}{{\frakm^\star}^\mu}\p^\mu\delf{\bar{F}}{{\frakm^\star}^\nu} - \delf{\bar{F}}{{\frakm^\star}^\mu}\p^\mu\delf{\bar{G}}{{\frakm^\star}^\nu} \ri) \nonumber\\
	&+& \sigma_\rmi\lf( \delf{\bar{G}}{{\frakm^\star}^\nu}\p^\nu\delf{\bar{F}}{\sigma_\rmi} - \delf{\bar{F}}{{\frakm^\star}^\nu}\p^\nu\delf{\bar{G}}{\sigma_\rmi} \ri) \nonumber\\
	\nonumber \\
	&+& \sigma_\rme\lf( \delf{\bar{G}}{{\frakm^\star}^\nu}\p^\nu\delf{\bar{F}}{\sigma_\rme} - \delf{\bar{F}}{{\frakm^\star}^\nu}\p^\nu\delf{\bar{G}}{\sigma_\rme} \ri) \nonumber\\
  &-& 2d_\rmi\lf( \delf{\bar{F}}{\sigma_\rme}\p^\mu\delf{\bar{G}}{{\calF^\star}^{\mu\nu}} - \delf{\bar{G}}{\sigma_\rme}\p^\mu\delf{\bar{F}}{{F^\star}^{\mu\nu}} \ri)\p^\nu\f{\sigma_\rme}{n} \nonumber\\
	\nonumber \\
	&+& 2\lf( \delf{\bar{F}}{{\frakm^\star}^\lambda}\p^\mu\delf{\bar{G}}{{\calF^\star}^{\mu\nu}} - \delf{\bar{G}}{{\frakm^\star}^\lambda}\p^\mu\delf{\bar{F}}{{\calF^\star}^{\mu\nu}} \ri){\calF^\star}^{\nu\lambda} \nonumber\\
  &+& \f{4d_\rmi}{n}\lf( \p^\mu\delf{\bar{F}}{{\calF^\star}^{\mu\nu}} \ri)\lf( \p^\lambda\delf{\bar{G}}{{\calF^\star}^{\lambda\kappa}} \ri){\calF^\star}^{\kappa\nu} \Bigg\}\rmd^4x. 
\label{e:XMHD noncanonical bracket electron-ion}
\end{eqnarray}
%-------------------------------------------------------
The bracket AP with this scaling gives the following equations:
%-------------------------------------------------------
\begin{eqnarray}
	&& \p_\nu\lf[ nhu^\mu u^\nu - d_\rmi h_\rme(u^\mu J^\nu + J^\mu u^\nu) + d_\rmi^2\f{h_\rme}{n}J^\mu J^\nu \ri] \nonumber\\
  && \hspace{12em} = \p^\mu p + J^\nu {\calF^\mu}_\nu, \label{e:XMHD e.o.m. electron-ion} \\
	&& \lf( u_\nu - d_\rmi\f{J_\nu}{n} \ri) {\calF^\star}^{\mu\nu} = -d_\rmi T_\rme\p^\mu\lf( \f{\sigma_\rme}{n} \ri) 
	\label{e:XMHD Ohm's law electron-ion} \\
	&& \p_\nu\lf(\sigma_\rmi u^\nu \ri) = 0, 
	\label{e:XMHD adiabaticity i electron-ion} \\
	&& \p_\nu\lf[\sigma_\rme \lf( u^\nu - d_\rmi\f{J^\nu}{n} \ri)\ri] = 0.
	\label{e:XMHD adiabaticity e electron-ion} 
\end{eqnarray}
%-------------------------------------------------------
Note,  this relativistic HMHD is  different from usual  nonrelativistic HMHD.  In Sec.~\ref{s:ColRecon} we explore some consequences of this.  

Next, upon taking  the limit $d_\rmi \to 0$, we obtain relativistic MHD.\cite{Lichnerowicz1967,Dixon1978,Anile1990} 
The Poisson bracket for the  relativistic MHD obtained by this reduction is different from the one proposed by D'Avignon \textit{et al.} in Ref.~\onlinecite{DAvignon2015} because a magnetic field like four vector $b^\nu$ was used there  instead of $A^{\mu}$.
The relation between the two brackets has yet to be clarified. 

The same reduction procedure (from XMHD to MHD) is applicable for the constrained least AP of Sec.~\ref{s:constrained least AP}.  
For example, if we ignore the electron rest mass, the velocities of each species are reduced as ${u_+}^\mu \to {u}^\mu$ and ${u_-}^\mu \to {u}^\mu - J^\mu/ne$. Similarly, the entropy and Lagrangian label constraints are reduced accordingly.
With these reductions the constrained least AP gives the relativistic HMHD equations.  We note, the  renormalization method used  in Ref.~\onlinecite{Yoshida2013} to derive AP for MHD  is also applicable for  relativistic HMHD. 

There are formalisms alternative to the one we presented that employ either Ohm's law or the induction equation \textit{per se} as a constraint  for nonrelativistic~\cite{Webb2014} and relativistic~\cite{Bekenstein2000,Bekenstein2001} MHD. 
However, these formulations cannot be reduced from the constrained action (\ref{e:XMHD S constrained}).
Whereas the physical meaning of the constraints in (\ref{e:XMHD S constrained}) is obvious, embedding the Ohm's law as a constraint is unnatural and arbitrary.
Furthermore, the EM field cannot be expressed by Clebsch potentials from  the AP with the Ohm's law constraint, unlike the case for our formulation where this emerges naturally  in  (\ref{e:XMHD delta J}).

\section{Relativistic Collisionless Reconnection}
\label{s:ColRecon} 

In nonrelativistic MHD with the inclusion of electron (rest mass) inertia (i.e., IMHD), a consequence of electron inertia  is  the violation of  the frozen-in magnetic flux condition, and instead, a flux determined by a generalized field is conserved.\cite{Lingam2015a}
Such an electron inertia effect was  suggested as a  mechanism for a collisionless magnetic reconnection\cite{Ottaviani1993} and has now been widely studied. 
However,  nonrelativistic HMHD does satisfy  the frozen-in magnetic flux  condition because the electron inertia is discarded by the $\mu \to 0$ limit.  Hence,  there is no direct mechanism causing  collisionless reconnection in nonrelativistic HMHD.

On the other hand, in relativistic XMHD, there are two kinds of electron inertiae: 
one from the electron rest mass $m_\rme$ and the other from the electron temperature $h_\rme$.
The $\mu \to 0$ limit corresponds to neglecting the former and keeping the latter.
Even though the former is small, the latter may not be ignorable when electron temperature is large enough.
The latter effect still allows for the violation of the frozen-in magnetic  flux condition.
Such a collisionless reconnection mechanism was previously proposed by Comisso~\textit{et al.}\  using a Sweet--Parker model in the context of  relativistic XMHD.\cite{Comisso2014}  
Here we find an  alternative flux given by the generalized vector potential ${A^\star}^\nu \to A^\nu - d_\rmi h_\rme u^\nu + d_\rmi^2(h_\rme/n)J^\nu$ to be  frozen-in.

Let us stress the difference between our  present study and the pair plasma study by Comisso~\textit{et al.}\cite{Comisso2014}
In the latter,  the relativistic electron--positron plasma with the assumption $\Delta h = 0$ was considered.
For HMHD, however, this $\Delta h = 0$ assumption removes the  aforementioned collisionless reconnection mechanism. 
From \eqref{e:Astar electron-ion} and \eqref{e:Adagger electron-ion}, we find ${A^\star}^\mu \to A^\mu$ and ${A^\dagger}^\mu \to A^\mu$ when we take both $\Delta h = 0$ and $\mu = 0$, so there is no longer the alternative frozen-in flux in HMHD. 

To make this statement more explicit, we write the relativistic HMHD induction equation in a  reference frame 
 moving with the  center-of-mass  (ion) velocity.  
 When the electron fluid is  homentropic, the right-hand side of (\ref{e:XMHD Ohm's law electron-ion}) vanishes.
Taking a curl of a spatial component of (\ref{e:XMHD Ohm's law electron-ion}), we obtain the induction equation in the reference frame, 
%-------------------------------------------------------
\begin{equation}
  \p_t\bm{B}^\star + \Curl(\bm{B}^\star\times\widetilde{\bm{v}}_\rme) = 0,
	\label{e:HMHD 3D induciton eq.}
\end{equation}
%-------------------------------------------------------
where 
%-------------------------------------------------------
\begin{equation}
	\bm{B}^\star = \bm{B} + \Curl\lf( -d_\rmi h_\rme\gamma\bm{v} + d_\rmi^2 h_\rme\f{\bm{J}}{n} \ri),
	\label{e:Bstar evolution}
\end{equation}
%-------------------------------------------------------
and
%-------------------------------------------------------
\begin{equation}
	\widetilde{\bm{v}}_\rme = \lf( \bm{v} - d_\rmi \f{\bm{J}}{\gamma n} \ri) \lf( 1 - d_\rmi\f{\rho_q}{\gamma n} \ri)^{-1}.
	\label{e:def vB}
\end{equation}
%-------------------------------------------------------
Here, $\widetilde{\bm{v}}_\rme$ is a modified electron velocity that becomes the electron velocity $\bm{v}_\rme$ in the  nonrelativistic limit $\gamma \to 1$ and $\rho_q \to 0$.
Evidently from \eqref{e:HMHD 3D induciton eq.} and \eqref{e:Bstar evolution}, the  magnetic field $\bm{B}$ is no longer frozen-in. 

Let us compare (\ref{e:Bstar evolution}) with the induction equations for other magnetohydrodynamic models,  summarized in Table~\ref{t:induction equations}.
The frozen-in condition for $\bm{B}$ is satisfied in nonrelativistic MHD, HMHD,  and relativistic MHD.
The frozen-in condition is violated in nonrelativistic two-dimensional IMHD,  while the alternative field $\bm{B} + \Curl(d_\rme^2\bm{J}/n)$, with the electron skin depth $d_\rme$ as characteristic length,\cite{Ottaviani1993} is frozen-in.
Therefore, the scale length of the collisionless reconnection caused by the electron inertia is $d_\rme$.
On the other hand, the alternative frozen-in field in relativistic HMHD is $\bm{B} + \Curl( -d_\rmi h_\rme\gamma\bm{v} + d_\rmi^2 h_\rme\bm{J}/n)$, which has a characteristic scale length with $\sqrt{h_\rme}d_\rmi$.
Since the scale length $d_\rme$ in  nonrelativistic IMHD is replaced to $\sqrt{h_\rme}d_\rmi$ in relativistic HMHD, and the reconnection scale is expected to be $\sqrt{h_\rme}d_\rmi$.  This estimate is the same as that for  the Sweet--Parker model for   relativistic electron--positron XMHD~\cite{Comisso2014} (recall that $h_\rme$ is normalized by $mc^2$ in this study).

Here we have inferred the reconnection scale just by comparing non-relativistic and relativistic Ohm's law.
However, in non-relativistic case, it was shown that the reconnection scale is not determined by the generalized Ohm's law alone when there is a strong magnetic guide field and appropriate gyro-physics is added to the model.
The analysis of the resulting gyrofluid model revealed that the relevant scale becomes the ion sound Larmor radius in this case.~\cite{Comisso2013}
Inclusion of gyroscopic effects in the relativistic context, appropriate for strong guide fields, is a subject for future work. 

%-------------------------------------------------------
\begin{table*}[htpb]
\caption{Induction equations for nonrelativistic MHD, HMHD and IMHD, and relativistic MHD and HMHD.}
\label{t:induction equations}
	\centering
	{\renewcommand{\arraystretch}{1.5}%
	{\setlength{\tabcolsep}{1.0em}
	\begin{tabular}{|r|c|c|}
		\hline
		& Barotropic induction eq. & frozen-in field \\
		\hline
		Nonrelativistic MHD  & $\p_t\bm{B} + \Curl(\bm{B}\times\bm{v}) = 0    $ & $\bm{B}$ \\
		\hline
		Nonrelativistic HMHD & $\p_t\bm{B} + \Curl(\bm{B}\times\bm{v}_\rme) = 0$ & $\bm{B}$ \\
		\hline
		Nonrelativistic 2D IMHD & $\p_t\bm{B}^\star + \Curl(\bm{B}^\star\times\bm{v}) = 0$ & $\ds \bm{B}^\star = \bm{B} + \Curl\lf(d_\rme^2\bm{J}/n \ri)$\\
		\hline
		Relativistic  MHD  & $\p_t\bm{B} + \Curl(\bm{B}\times\bm{v}) = 0     $ & $\bm{B}$ \\
		\hline
		Relativistic  HMHD & $\p_t\bm{B}^\star + \Curl(\bm{B}^\star\times\widetilde{\bm{v}}_\rme) = 0 $ & $\ds \bm{B}^\star = \bm{B} + \Curl\lf( -d_\rmi h_\rme\gamma\bm{v} + d_\rmi^2 h_\rme\bm{J}/n \ri)$ \\
		\hline
	\end{tabular}
	}
	}
\end{table*}
%-------------------------------------------------------

%-------------------------------------------------------
% Limit to nonrelativistic XMHD
%-------------------------------------------------------
\section{Nonrelativistic XMHD -- 3+1 decomposition}
\label{s:NR limit}

The covariant Poisson bracket AP formalism also encompasses nonrelativistic theories.   We will show this in the context of XMHD, then infer that this is the case for nonrelativistic MHD and the nonrelativistic ideal fluid.  
Because nonrelativistic theories contain space and time separately, it is natural to pursue this end by beginning from the 3+1 decomposition for relativistic theories described in Ref.~\onlinecite{Marsden1986}.  To this end we state some general tools before proceeding to the task at hand.

The functional derivative of  \eqref{e:4FD} is defined relative to the space-time pairing, while functional derivatives in conventional Hamiltonian theories are defined relative to only  the spatial pairing, i.e. 
\begin{equation}
\delta \mathscr{F}=\int \delta n\, \frac{\delta \mathscr{F}}{\delta n}\, \rmd^3 x\,.
\label{e:3FD}
\end{equation}
For functionals of the form $F=\int  \mathscr{F} \rmd x^0$ where $\mathscr{F}$ contains no time derivatives of a field, it follows e.g.\  that 
\begin{equation}
\frac{\delta F}{\delta n(x^0,\bm{x})}=    \frac{\delta \mathscr{F}}{\delta n(\bm{x})}\,,
\end{equation}
where we explicitly display the arguments to distinguish space-time  from space functional derivatives.   For nonrelativistic theories, we need to consider functionals that are localized in time, i.e., have the form 
\begin{equation}
F=\int  \delta(x^0-{x^0}')\, \mathscr{F} \,  \rmd x^0\,. 
\label{e:locF}
\end{equation}
Observe,  in this case, if $\mathscr{F}$ contains no time derivatives of the field  $n$, then  
\begin{equation}
\frac{\delta F}{\delta n(x^0,\bm{x})}=  \delta(x^0-{x^0}')\,  \frac{\delta \mathscr{F}}{\delta n(\bm{x})}\,,
\end{equation}
and similarly for other fields. 
Next, let us suppose that a functional $G$ is separable in the following sense
\begin{equation}
G= G^0 + \int  \mathscr{G} \,  \rmd x^0\,,
\label{e:sepG}
\end{equation}
where all of the time-like components of fields are contained in the functionals $G^0$ and $\mathscr{G}$ contains no time derivatives of fields.   For functionals $G$ of the form of \eqref{e:sepG}  and $F$ of the form of \eqref{e:locF}, it will be shown that 
\begin{equation}
0=\{F,G\}= - \frac{\rmd \mathscr{F}}{\rmd t} + \{ \mathscr{F}, \mathscr{G}\}^{(3)}
\end{equation}
where $\{ F, G \}$ is the canonical bracket \eqref{e:XMHD canonical bracket} or the noncanonical bracket \eqref{e:XMHD noncanonical bracket}, and $ \{ \mathscr{F}, \mathscr{G}\}^{(3)}$ is the appropriate nonrelativistic Poisson bracket.  In this way one can establish the connection between Poisson bracket APs and usual noncanonical Poisson bracket Hamiltonian formulations. 

For the case at hand, let us return to the arbitrary mass plasma ($m_+$ and $m_-$) and consider a nonrelativistic limit with
%-------------------------------------------------------
\begin{equation*}
  h \to mc^2, \;\; \Delta h \to 0, \;\; h^\dagger \to (m_+ m_-)c^2/m, \;\; \gamma_\pm \to 1, \; \p_t \bm{E} = 0.
\end{equation*}
%-------------------------------------------------------
These result in $J^0 = en(\gamma_+ - \gamma_-) \to 0$ and $\bm{J} = \Curl\bm{B}$, and the 
 generalized fields become
%-------------------------------------------------------
\begin{eqnarray*}
  {\frakm^\star}^\nu &\to& nmc^2 u^\nu = \frakm^\nu\,,\\
   {A^\star}^\nu &\to& A^\nu + \f{1}{ne^2}\lf( \f{m_+ m_-}{m}c^2 \ri)J^\nu\,,
   \end{eqnarray*}
   and 
   \[
 {A^\dagger}^\nu = \f{m_+ - m_-}{m}{A^\star}^\nu - \f{m_- m_+ c}{m e}u^\nu\,,
\]
%-------------------------------------------------------
with the four-velocity becoming  $u^\nu = (1,\, \bm{v}/c)$.
Using the thermodynamic relations $\rho = n(mc^2 + \calE)$ and $p = nh - \rho$, with internal energy $\calE$, the following limit is calculated
%-------------------------------------------------------
\begin{eqnarray*}
  \f{1}{2}(\rho - p) = n(mc^2 + \calE) - \f{1}{2}nh \; \to \; n\lf( \f{1}{2}mc^2 + \calE \ri).
\end{eqnarray*}
%-------------------------------------------------------

We first show a nonrelativistic Hamilton's equation for the Clebsch variables.
The action \eqref{e:XMHD S constrained} is separated as
\begin{eqnarray*}
  S[z] = \int \lf[ n\p^0\phi + \sum_\pm \lf( \sigma_\pm\p^0\eta_\pm + \varphi_\pm\p^0\lambda_\pm \ri) \ri]\,\rmd^4x - \int \scrH\rmd^0x,
\label{e:NR XMHD action Clebsch}
\end{eqnarray*}
with a Hamiltonian
%-------------------------------------------------------
\begin{eqnarray}
  \scrH\lf[ z \ri] = \int \lf[ n(mc^2 +\calE_+ + \calE_-) + \f{1}{2}nmv^2 + \f{1}{2}\bm{J}\cdot\bm{A}^\star \ri] \rmd^3 x. \nonumber\\
\end{eqnarray}
%-------------------------------------------------------
Here $\bm{v} = \bm{\frakm}/nmc$, $\bm{A}^\star$, and $\bm{J}$ are functions of $z$.
Substituting, this action and the localized functional \eqref{e:locF} into the covariant canonical bracket \eqref{e:XMHD canonical bracket}, we get
%-------------------------------------------------------
\begin{eqnarray*}
  \lf\{ F,\, S \ri\}_\mr{canonical} = \int \lf( \dd{\scrF}{t} - \lf\{ \scrF,\, \scrH \ri\}^{(3)}_\mathrm{canonical} \ri)\delta(x^0-x^{0'})\,\rmd x^0
\end{eqnarray*}
%-------------------------------------------------------
where $\lf\{ \scrF,\, \scrH \ri\}^{(3)}_\mathrm{canonical}$ is a canonical Poisson bracket defined in the three-dimensional space.
Thus, we get the nonrelativistic canonical Hamilton's equation as
%-------------------------------------------------------
\begin{eqnarray*}
  \dd{\scrF}{t} = \lf\{ \scrF,\, \scrH \ri\}^{(3)}_\mathrm{canonical},
\end{eqnarray*}
%-------------------------------------------------------
which describes the time evolution of the Clebsch variables $z$.
Transforming the Clebsch variables to $\bm{v}$ and $\bm{B}^\star$, we obtain the non-relativistic XMHD equations, which will be explicitly shown below.

Now we are set to apply this 3+1 procedure to the noncanonical bracket \eqref{e:XMHD noncanonical bracket}.
Upon rearranging the action of (\ref{e:XMHD S}) we obtain 
%-------------------------------------------------------
\begin{eqnarray}
  \bar{S} = \int \f{\frakm_0 \frakm^0}{2nmc^2}\rmd^4x - \int\scrH\rmd x^0,
\label{e:NR XMHD action}
\end{eqnarray}
%-------------------------------------------------------
with the  Hamiltonian
%-------------------------------------------------------
\begin{eqnarray}
  && \scrH\lf[ n,\, \sigma,\, \frakm^i,\, {A^\star}^i \ri] =
  \label{e:NR XMHD Hamiltonian}\\ 
  && \int \lf[ -\f{\frakm_i \frakm^i}{2nmc^2} + n\lf( \f{1}{2}mc^2 + \calE_+ + \calE_- \ri) - \f{{A^\star}^i J_i}{2} \ri] \rmd^3 x,
 \nonumber
\end{eqnarray}
%-------------------------------------------------------
where we used $J^0 = 0$ to get the last term.

Then we calculate $\{ \bar{F},\, \bar{S} \}_\mathrm{XMHD} = 0$ to get the  nonrelativistic XMHD equations.
The phase space variables must be $(n,\, \sigma_\pm,\, \frakm^i, {A^\star}^i)$.
Hence we put $\bar{F} = \bar{F}[n,\, \sigma_\pm,\, \frakm^i, {A^\star}^i]$.
Since the action (\ref{e:NR XMHD action}) does not depends on ${A^\star}^0$, we may write $\bar{S} = \bar{S}[n,\, \sigma_\pm,\, \frakm^0,\, \frakm^i, {A^\star}^i]$.
Therefore all the terms including $\delta \bar{F}/\delta \frakm^0$, $\delta \bar{F}/\delta {A^\star}^0$, and $\delta \bar{S}/\delta {A^\star}^0$ are dropped.
Upon writing 
%-------------------------------------------------------
\begin{eqnarray*}
  \bar{F} = \int \delta(x^0-{x^0}')\,\scrF[n,\, \sigma_\pm,\, \frakm^i, {A^\star}^i]\,\rmd x^0\,, 
\end{eqnarray*}
%-------------------------------------------------------
 the covariant bracket AP can be  written as
%-------------------------------------------------------
% \begin{widetext}
% \begin{eqnarray*}
  % 0=\lf\{ \bar{F},\, \bar{S} \ri\} &=& \int\Bigg\{ n\f{\frakm_0}{nmc^2}\p^0 \de{\scrF}{n} + n\lf[ -\de{\scrH}{\frakm^i}\p^i\de{\scrF}{n} + \de{\scrF}{\frakm^i}\p^i\lf( \f{\frakm_0 \frakm^0}{2n^2mc^2} + \de{\scrH}{n} \ri) \ri] \nonumber\\
  % \\
  % & & \hspace{1em} + \sum_\pm \Bigg[ \sigma_\pm\f{\frakm_0}{nmc^2}\p^0 \de{\scrF}{\sigma_\pm} + \sigma_\pm\lf[ -\de{\scrH}{\frakm^i}\p^i\de{\scrF}{\sigma_\pm} + \de{\scrF}{\frakm^i}\p^i\lf( \de{\scrH}{\sigma_\pm} \ri) \ri] \pm \f{m_\mp}{me}\lf( \delf{\scrF}{\sigma_\pm}\delf{\scrH}{{A^\star}^{i}} - \delf{\scrH}{\sigma_\pm}\delf{\scrF}{{A^\star}^{i}} \ri)\p^i\lf( \f{\sigma_\pm}{n} \ri) \Bigg]\nonumber\\
  % \\
  % & & \hspace{1em} + \frakm^i\lf( \f{\frakm^0}{nmc^2}\p^0\de{\scrF}{\frakm^i} \ri) +  \frakm^0\lf( -\de{\scrF}{\frakm^i}\p^i\f{\frakm^0}{nmc^2} \ri) - {\frakm}^j\lf( \de{\scrH}{{\frakm}^i}\p^i\de{\scrF}{{\frakm}^j} - \de{\scrF}{{\frakm}^i}\p^i\de{\scrH}{{\frakm}^j} \ri) \\
  % & & \hspace{1em} + \lf( \f{\frakm_0}{nmc^2}\delf{\scrF}{{A^\star}^{i}} \ri)\lf( \p^i{A^\star}^0 - \p^0{A^\star}^i \ri) \textcolor{red}{+ \lf( \delf{\scrH}{{\frakm}^i}\delf{\scrF}{{A^\star}^{j}} - \delf{\scrF}{{\frakm}^i}\delf{\scrH}{{A^\star}^{j}} \ri){F^\star}^{ji} - \f{1}{ne}\delf{\scrF}{{A^\star}^{i}}\delf{\scrF}{{A^\star}^{j}}{F^\dagger}^{ij}} \Bigg\}\zeta(t)\;\rmd^4x.
% \end{eqnarray*}
% \end{widetext}
\begin{eqnarray*}
  0&=&\lf\{ \bar{F},\, \bar{S} \ri\}_\mathrm{XMHD}
  = -\int\Bigg\{ n\f{\frakm_0}{nmc^2}\p^0 \lf[\delta(x^0-{x^0}')\,\de{\scrF}{n}\ri] \\
  &+&\delta(x^0-{x^0}')\, n\lf[ -\de{\scrH}{\frakm^i}\p^i\de{\scrF}{n} + \de{\scrF}{\frakm^i}\p^i\lf( \f{\frakm_0 \frakm^0}{2n^2mc^2} + \de{\scrH}{n} \ri) \ri] \nonumber\\
  &+& \sum_\pm \Bigg( \sigma_\pm\f{\frakm_0}{nmc^2}\p^0 \lf[\delta(x^0-{x^0}')\, \de{\scrF}{\sigma_\pm} \ri] \\
  & & \hspace{1em} +\delta(x^0-{x^0}')\,\sigma_\pm\lf[ -\de{\scrH}{\frakm^i}\p^i\de{\scrF}{\sigma_\pm} + \de{\scrF}{\frakm^i}\p^i\lf( \de{\scrH}{\sigma_\pm} \ri) \ri] \\
  & & \hspace{1em} \pm \delta(x^0-{x^0}')\,\f{m_\mp}{me}\lf( \delf{\scrF}{\sigma_\pm}\delf{\scrH}{{A^\star}^{i}} - \delf{\scrH}{\sigma_\pm}\delf{\scrF}{{A^\star}^{i}} \ri)\p^i\lf( \f{\sigma_\pm}{n} \ri) \Bigg)\nonumber\\
  &+& \frakm^i\lf( \f{\frakm^0}{nmc^2}\p^0\lf[\delta(x^0-{x^0}')\,\de{\scrF}{\frakm^i}\ri] \ri) +  \delta(x^0-{x^0}')\,\frakm^0\lf( -\de{\scrF}{\frakm^i}\p^i\f{\frakm^0}{nmc^2} \ri) \\
  &-&\delta(x^0-{x^0}')\, {\frakm}^j\lf( \de{\scrH}{{\frakm}^i}\p^i\de{\scrF}{{\frakm}^j} - \de{\scrF}{{\frakm}^i}\p^i\de{\scrH}{{\frakm}^j} \ri) \\
  &+& \delta(x^0-{x^0}')\,\lf( \f{\frakm_0}{nmc^2}\delf{\scrF}{{A^\star}^{i}} \ri)\lf( \p^i{A^\star}^0 - \p^0{A^\star}^i \ri) \\
  &+&\delta(x^0-{x^0}')\,\lf( \delf{\scrH}{{\frakm}^i}\delf{\scrF}{{A^\star}^{j}} - \delf{\scrF}{{\frakm}^i}\delf{\scrH}{{A^\star}^{j}} \ri){F^\star}^{ji} \\
  &-&\delta(x^0-{x^0}')\, \f{1}{ne}\delf{\scrF}{{A^\star}^{i}}\delf{\scrF}{{A^\star}^{j}}{F^\dagger}^{ij} \Bigg\}\;\rmd^4x.
\end{eqnarray*}
%-------------------------------------------------------
Next  we substitute $\frakm^0 = \frakm_0 = nmc^2$ and  manipulate some of the terms to obtain 
%-------------------------------------------------------
\begin{eqnarray*}
  && \int \Bigg( \delf{\scrF}{n}\p^0 n + \delf{\scrF}{\sigma_+}\p^0 \sigma_+ + \delf{\scrF}{\sigma_-}\p^0 \sigma_- \\
  && \hspace{5em} + \delf{\scrF}{\frakm^i}\p^0 \frakm^i + \delf{\scrF}{{A^\star}^i}\p^0 {A^\star}^i \Bigg) \,\rmd^3x = \f{1}{c}\dd{\scrF}{t}\,, 
\end{eqnarray*}
%-------------------------------------------------------
yielding 
%-------------------------------------------------------
\begin{eqnarray*}
  && \lf\{ \bar{F},\, \bar{S} \ri\}_\mathrm{XMHD} = \f{1}{c}\int \lf( \dd{\scrF}{t} - \lf\{ \scrF,\, \scrH \ri\}^{(3)} \ri)\delta(x^0-{x^0}')\,\rmd x^0\\
  && \hspace{8em} + \int{A^\star}^0\p^i\lf( \delf{\scrF}{{A^\star}^{i}} \ri)\delta(x^0-{x^0}')\,\rmd^4x\,,
\end{eqnarray*}
%-------------------------------------------------------
with a three dimensional Poisson bracket $\lf\{ \scrF,\, \scrG \ri\}^{(3)}$ that will be explicitly shown below.
Evaluating the $\delta$-function shows $\lf\{ \bar{F}, \bar{S} \ri\}_\mathrm{XMHD} = 0$ is equivalent to Hamilton's equation along with a gauge-like condition: 
%-------------------------------------------------------
\begin{eqnarray}
  \dd{\scrF}{t} = \lf\{ \scrF,\, \scrH \ri\}^{(3)} \quad \text{and} \quad \Div\lf( \delf{\scrF}{{\bm{A}^\star}} \ri) = 0.
  \label{e:NR XMHD Hamiltonian eq.}
\end{eqnarray}
%-------------------------------------------------------

The second equation of  (\ref{e:NR XMHD Hamiltonian eq.}), the gauge condition,  is handled  manifestly by transforming from  the phase space  variable $\bm{A}^\star$ to $\bm{B}^\star$; since $\delta \scrF/\delta \bm{A}^\star = \Curl(\delta \scrF/\delta \bm{B}^\star)$,  with this transformation the second condition is automatically satisfied.
Finally, we  transform $\bm{\frakm}$ to $\bm{v}$ and find that  the Poisson bracket $\lf\{ \scrF,\, \scrG \ri\}^{(3)}$ becomes 
%------------------------------------------------------
% \begin{widetext}
% \begin{eqnarray}
  % \lf\{ \scrF,\, \scrG \ri\}^{(3)} &=& \int\lf\{ \lf( \de{\scrG}{{\bm{v}}}\cdot\nbl\de{\scrF}{\varrho} - \de{\scrF}{{\bm{v}}}\cdot\nbl\de{\scrG}{\varrho} \ri) + \f{\Curl\bm{v}}{\varrho}\cdot\lf( \de{\scrF}{\bm{v}}\times\de{\scrG}{\bm{v}} \ri) \ri. \nonumber\\
	% \nonumber \\
  % & & \hspace{1em} + \sum_\pm \lf[ \f{\sigma_\pm}{\varrho}\lf( \de{\scrG}{{\bm{v}}}\cdot\nbl\de{\scrF}{\sigma_\pm} - \de{\scrF}{{\bm{v}}}\cdot\nbl\de{\scrG}{\sigma_\pm} \ri) \mp \f{cm_\mp}{e}\lf( \de{\scrF}{\sigma_\pm}\lf( \Curl\delf{\scrG}{{\bm{B}^\star}} \ri) - \de{\scrG}{\sigma_\pm}\lf( \Curl\delf{\scrF}{{\bm{B}^\star}} \ri) \ri)\cdot\nbl\lf( \f{\sigma_\pm}{\varrho} \ri) \ri] \nonumber\\
  % & & \hspace{1em} \lf. - \lf[ \de{\scrG}{{\bm{v}}}\times\lf( \Curl\delf{\scrF}{{\bm{B}^\star}} \ri) - \de{\scrF}{{\bm{v}}}\times\lf( \Curl\delf{\scrG}{{\bm{B}^\star}} \ri) \ri]\cdot\f{\bm{B}^\star}{\varrho} - \textcolor{red}{\f{mc}{\varrho e}\lf[ \lf( \Curl\delf{\scrF}{{\bm{B}^\star}} \ri)\times\lf( \Curl\delf{\scrG}{{\bm{B}^\star}} \ri) \ri]\cdot{\bm{B}^\dagger}} \ri\}\rmd^3x, 
  % \label{e:NR XMHD Bracket Bstar and v}
% \end{eqnarray}
% \end{widetext}
\begin{eqnarray}
  \lf\{ \scrF,\, \scrG \ri\}^{(3)} &=& \int\Bigg\{ \lf( \de{\scrG}{{\bm{v}}}\cdot\nbl\de{\scrF}{\varrho} - \de{\scrF}{{\bm{v}}}\cdot\nbl\de{\scrG}{\varrho} \ri) \nonumber\\
  &+& \f{\Curl\bm{v}}{\varrho}\cdot\lf( \de{\scrF}{\bm{v}}\times\de{\scrG}{\bm{v}} \ri) \nonumber\\
  &+& \sum_\pm \Bigg[ \f{\sigma_\pm}{\varrho}\lf( \de{\scrG}{{\bm{v}}}\cdot\nbl\de{\scrF}{\sigma_\pm} - \de{\scrF}{{\bm{v}}}\cdot\nbl\de{\scrG}{\sigma_\pm} \ri) \nonumber\\
  & &\mp \f{cm_\mp}{e}\Bigg( \de{\scrF}{\sigma_\pm}\lf( \Curl\delf{\scrG}{{\bm{B}^\star}} \ri) \nonumber\\
  & & \hspace{6em}- \de{\scrG}{\sigma_\pm}\lf( \Curl\delf{\scrF}{{\bm{B}^\star}} \ri) \Bigg)\cdot\nbl\lf( \f{\sigma_\pm}{\varrho} \ri) \Bigg] \nonumber\\
  &-& \lf[ \de{\scrG}{{\bm{v}}}\times\lf( \Curl\delf{\scrF}{{\bm{B}^\star}} \ri) - \de{\scrF}{{\bm{v}}}\times\lf( \Curl\delf{\scrG}{{\bm{B}^\star}} \ri) \ri]\cdot\f{\bm{B}^\star}{\varrho} \nonumber\\
  &-& \f{mc}{\varrho e}\lf[ \lf( \Curl\delf{\scrF}{{\bm{B}^\star}} \ri)\times\lf( \Curl\delf{\scrG}{{\bm{B}^\star}} \ri) \ri]\cdot{\bm{B}^\dagger} \Bigg\}\,\rmd^3x, 
  \label{e:NR XMHD Bracket Bstar and v}
\end{eqnarray}
%-------------------------------------------------------
where $\varrho = mn$ and $\nbl = -\p^i$.
This Poisson bracket is a generalization of the nonrelativistic electron-ion XMHD bracket proposed before~\cite{Abdelhamid2015,Lingam2015}. The  bracket of \eqref{e:NR XMHD Bracket Bstar and v} differs from the previous results by the choice of scaling and,  more importantly,  the assumption  $m_-\ll m_+$ is not made.

Now consider the  Hamiltonian  of (\ref{e:NR XMHD Hamiltonian});   it  becomes
%-------------------------------------------------------
\begin{eqnarray}
  && \scrH\lf[ n,\, \sigma,\, \bm{v},\, \bm{B}^\star \ri] = \nonumber\\
  && \hspace{1em} \int \lf[ \f{\varrho|\bm{v}|^2}{2} + \varrho\lf( \f{1}{2}mc^2 + \calE_+ + \calE_- \ri) + \f{\bm{B}^\star\cdot\bm{B}}{2} \ri] \rmd^3 x,
\label{e:NR XMHD Hamiltonian Bstar and v}
\end{eqnarray}
%-------------------------------------------------------
where $\calE_{\pm}/m$ is rewritten as $\calE_{\pm}$.
The functional derivatives of $\scrH$ are
%-------------------------------------------------------
\begin{eqnarray*}
  &&\de{\scrH}{\varrho} = \f{1}{m}\de{\scrH}{n} = \f{v^2}{2} + \f{c^2}{2} + \sum_\pm \lf( \calE_\pm + \varrho\pp{\calE_\pm}{\varrho} \ri) + \f{m_+ m_-c^2}{2\varrho^2e^2}J^2, \\
  &&\de{\scrH}{\sigma_\pm} = \varrho\pp{\calE_\pm}{\sigma_\pm}, \quad \de{\scrH}{\bm{v}} = \varrho\bm{v}, \quad \delf{\scrH}{\bm{B}^\star} = \bm{B}\,.
\end{eqnarray*}
%-------------------------------------------------------
Finally, using the above  Hamilton's equations of (\ref{e:NR XMHD Hamiltonian eq.}) give 
%-------------------------------------------------------
\begin{eqnarray*}
  \pp{\varrho}{t} &=& \lf\{ \varrho,\, \scrH \ri\}^{(3)} = -\Div(\varrho\bm{v}) \\
  \nonumber\\
  \pp{\sigma_\pm}{t} &=& \lf\{ \sigma_\pm,\, \scrH \ri\}^{(3)} = -\Div\lf[ \lf( \bm{v} \pm \f{cm_\mp}{\varrho e}\bm{J} \ri)\sigma_\pm \ri] \\
  \nonumber\\
  \pp{\bm{B}^\star}{t} &=& \lf\{ \bm{B}^\star,\, \scrH \ri\}^{(3)} = \sum\pm\Curl\lf[ \f{cm_\mp}{e}T_\pm\nbl\lf( \f{\sigma_\mp}{\varrho} \ri) \ri] \nonumber \\
  &&\hspace{6em} + \Curl\lf( \bm{v}\times\bm{B}^\star \ri) - \Curl\lf( \f{mc}{\varrho e}\bm{J}\times\bm{B}^\dagger \ri) \\
  \nonumber\\
  \pp{\bm{v}}{t} &=& \lf\{ \bm{v},\, \scrH \ri\}^{(3)} = -(\Curl\bm{v})\times\bm{v} - \nbl\lf( \f{v^2}{2} + \f{m_+ m_-c^2}{2\varrho^2e^2}J^2 \ri) \nonumber \\
  &&\hspace{14em} - \f{\nbl p}{\varrho} + \f{\bm{J}\times\bm{B}^\star}{\varrho}\,, 
\end{eqnarray*}
%-------------------------------------------------------
 the nonrelativistic L\"{u}st equations~\cite{Lust1959}.  Note, here  we used the thermodynamic relations
%-------------------------------------------------------
\begin{eqnarray*}
  \rmd \calE = T\rmd \lf( \f{\sigma}{\varrho} \ri) + \f{p}{\varrho^2}\rmd \varrho = \f{T}{\varrho}\rmd \sigma + \f{1}{\varrho^2}(p - T\sigma)\rmd \varrho
\end{eqnarray*}
and
\begin{eqnarray*}
  \varrho\rmd\lf( \calE + \varrho\pp{\calE}{\varrho} \ri) + \sigma\rmd\lf( \varrho\pp{\calE}{\sigma} \ri) = \rmd p.
\end{eqnarray*}

In closing this section, we  seek the Casimirs of \eqref{e:NR XMHD Bracket Bstar and v} for the barotropic case.   They must satisfy $\forall F:\; 0 = \{F,C\}$ leading to a system
\begin{equation}\label{e:first casimir eq}
	\nabla\times\bigg(\f{\bm{B}^\star}{\varrho}\times C_{\bm{v}}+C_{\bm{A}^\star}\times\f{\bm{B}^\star}{\varrho}\bigg) = 0
\end{equation}
\begin{equation}\label{e:second casimir eq}
	\nabla\cdot C_{\bm{v}} = 0\quad\text{and}\quad C_{\bm{v}}\times\f{\nabla\times\bm{v}}{\varrho}+C_{\bm{A}^\star}\times\f{\bm{B}^\star}{\varrho}-\nabla C_\varrho = 0,
\end{equation}
where we use the abbreviated notation  $C_\varrho:=\delta C/\delta \varrho$.  Seeking a helicity  Casimir we assume   a linear combination
\begin{equation}
	C^{(\lambda)} = \f{1}{2} \int d^3 x\; \Big(\bm{A}^\star+\lambda\,\bm{v}\Big)\cdot\Big(\bm{B}^\star+\lambda\,\Curl\bm{v}\Big)\,, 
\end{equation}
which is substituted  into \eqref{e:first casimir eq} and \eqref{e:second casimir eq} leading  to a quadratic equation for $\lambda$ with roots $\lambda_\pm = \pm m_\pm c/e$. These new Casimirs constitute topological constraints for a plasma with $m_+$, $m_-$ species masses. In the limit $m_-\ll m_+$ these Casimirs become those of  Refs.~\onlinecite{Abdelhamid2015,Lingam2015}. For a discussion of topological properties of XMHD see Ref.~\cite{LMM16}. Notice that the $C^{\pm}$ coincide exactly with the known 2-fluid canonical helicities $\int P \wedge dP$ for each species  of Refs.~\onlinecite{Mahajan2003,Yoshida2014}. However we emphasize here  the importance of the variables $\bm v$ and $\bm{A}^\star$. 

In addition the helicity Casimirs, when barotropic condition is violated we obtain the family
	\begin{equation}
		C^{(\sigma)} = \int \rmd^3 x\; \varrho\, f\,\Big(\f{\sigma_+}{\varrho},\f{\sigma_-}{\varrho}\Big)\,,
	\end{equation}
albeit with the condition $\sigma_+/\varrho$ being a  function of $\sigma_-/\varrho$ or $f_{,+-} = 0$, where $f_{,+}$ denotes differentiation with respect to the first argument.

%-------------------------------------------------------

%-------------------------------------------------------
% Conclusion
%-------------------------------------------------------
\section{Conclusion}
\label{s:conclusion}
We have formulated  APs  for   relativistic XMHD.
For the constrained least action principle, the constraints, namely,   conservation of  number density, entropy, and   Lagrangian labels for each species, were employed in the manner of Lin.
Extremization of the constrained action led to Clebsch potential expressions for the generalized momentum   and the generalized vector potential. 
Then,  variable transformation from the Clebsch potentials to the physical variables led to the covariant Poisson bracket for XMHD.  In the Poisson bracket AP  the  constraints are hidden in the degeneracy of the Poisson bracket.   Through these APs we have unified the  Eulerian APs for all magnetofluid  models. Indeed, returning to  Table~\ref{t:knowns and unknowns} we see that all slots for Eulerian APs have been completed.  Now,  the only remaining work is the formulation of  the AP for relativistic XMHD in the Lagrangian description.  Examination of the results of Ref.~\onlinecite{DAvignon2016} for  nonrelativistic XMHD suggests this may not be an easy task. 
 
Another important result was our formulation of  relativistic HMHD, obtained by taking  a limit of  the AP for XMHD.
We observed that while nonrelativistic HMHD does not have a direct mechanism for collisionless reconnection,  relativistic HMHD does  allow the  violation of the frozen-in magnetic flux  condition via the electron  thermal inertia effect.
We also found  an  alternative frozen-in flux, in a manner similar to that for  nonrelativistic IMHD.
The scale length of the collisionless reconnection was shown to correspond  to the reconnection layer width estimated by the Sweet--Parker model.\cite{Comisso2014}  Further study of  relativistic HMHD, such as a numerical simulation of (\ref{e:HMHD 3D induciton eq.}), will be the subject of future work. 

Lastly  in this paper,  we passed  to a nonrelativistic limit within the covariant bracket formalism, thus arriving at a ``covariant'' bracket for nonrelativistic XMHD. Then we derived  the usual 3+1 noncanonical Poisson bracket. However, beyond  the results of Refs.~\onlinecite{Abdelhamid2015,LMM16},  the result of  \eqref{e:NR XMHD Bracket Bstar and v} does not assume smallness of electron mass and thus is also applicable to electron--positron plasmas.

\acknowledgments
The work of YK was supported by JSPS KAKENHI Grant Number 26800279. GM and PJM  were supported by U.S.\ Dept.\ 
of Energy under contract \#DE-FG02-04ER-54742. PJM  would also like to acknowledge support from the Alexander
von Humboldt Foundation and the hospitality of the Numerical Plasma Physics Division of the IPP, Max Planck, Garching.

\bibliographystyle{apsrev4-1}
\bibliography{references}

\end{document}